\documentclass[3p,twocolumn]{elsarticle}

\usepackage{lineno}
\usepackage{hyperref}
\usepackage{enumitem}
\modulolinenumbers[5]

\usepackage{amssymb}
\usepackage{amsmath}

\newcommand{\ket}[1]{\ensuremath{|#1\rangle}}
\newcommand{\bra}[1]{\ensuremath{\langle#1|}}
\newcommand{\inner}[2]{\ensuremath{\langle#1|#2\rangle}}
\newcommand{\matrixelem}[3]{\ensuremath{\langle#1|#2|#3\rangle}}

\newcommand{\nuc}[2]{\ensuremath{{}^{#1}\rm{#2}}}
\newcommand{\NOT}{\textsc{not}}

\newcommand{\CNOT}{controlled-\NOT}

\newcommand{\half}{\ensuremath{\textstyle\frac{1}{2}}}

\newcommand{\rmd}{\mathrm{d}}
\newcommand{\rme}{\mathrm{e}}
\newcommand{\rmi}{\mathrm{i}}

\journal{Progress in Nuclear Magnetic Resonance Spectroscopy}

\begin{document}

\begin{frontmatter}

\title{Controlling NMR spin systems for quantum computation}

\author{Jonathan A. Jones}
\address{Clarendon Laboratory, University of Oxford, Parks Road, Oxford OX1 3PU, UK}

\begin{abstract}
Nuclear magnetic resonance is arguably both the best available quantum technology for implementing simple quantum computing experiments and the worst technology for building large scale quantum computers that has ever been seriously put forward. After a few years of rapid growth, leading to an implementation of Shor’s quantum factoring algorithm in a seven-spin system, the field started to reach its natural limits and further progress became challenging. Rather than pursuing more complex algorithms on larger systems, interest has now largely moved into developing techniques for the precise and efficient manipulation of spin states with the aim of developing methods that can be applied in other more scalable technologies and within conventional NMR. However, the user friendliness of NMR implementations means that they remain popular for proof-of-principle demonstrations of simple quantum information protocols.
\end{abstract}


\end{frontmatter}
\tableofcontents

\section{Introduction}
Quantum information processing (QIP) is the use of explicitly quantum mechanical systems, exhibiting phenomena such as superposition and entanglement, to perform information processing tasks. Traditionally the field can be divided into two broad areas: quantum computation is about the performance of computational tasks more efficiently than is possible for any classical computer \cite{Deutsch1985}, while quantum communication largely considers tasks which are simply impossible by purely classical means \cite{Wiesner1983}. Closely related to quantum computation is quantum simulation, in which one quantum mechanical system is used to model, and thus study, the behaviour of another \cite{Feynman1982}. The distinction between computation and simulation is not always simple or clear \cite{Toffoli1982}, and the design of general-purpose quantum simulators is an active area \cite{Daley2022}. Another growing area is quantum sensing, in which non-classical states of light or atoms are used to achieve a measurement precision beyond the standard quantum limit \cite{Kimble1987}.

Over the last forty years QIP, and particularly quantum computing, has moved from a purely theoretical domain explored only by a few committed enthusiasts to a thoroughly mainstream area of science \cite{Bennett2000, Gisin2002, Kimble2008, NCbook, Merminbook, JonesJakschBook, Hidari2021, FlarendBook, Castelvecchi2022}. NMR experiments have played a small but significant role in this: early discussions of how NMR quantum computers could be implemented \cite{Cory1996, Cory1997, Cory1998a, Gershenfeld1997, Chuang1998b} were soon followed by the first implementations of complete quantum algorithms \cite{Jones1998c, Chuang1998, Chuang1998a, Jones1998d}. Indeed for a few years NMR was in many ways the leading quantum computation technology, culminating in the first implementation of Shor's quantum factoring algorithm \cite{Vandersypen2001}. This rapid progress was, however, matched by a corresponding concern: the difficulty of preparing NMR spin systems in pure states, a consequence of the tiny energy gap for nuclear spin levels, almost rules out attempts to build large scale devices \cite{Warren1997, Gershenfeld1997a}. Even if this were resolved many issues would remain, such as the difficulty of designing spin systems with very large networks of coupled spins which permit sufficiently selective excitation \cite{Jones2000a}. For these reasons NMR quantum computing has been described as a demonstration technology \cite{Jones1999b}, and as a field for developing tricks and techniques which will find their final applications in other fields \cite{Jones2001}.

The role of NMR in studies of quantum communication has been even more limited for two basic reasons. Simple quantum communication protocols, such as BB84 quantum cryptography \cite{Bennett1992b}, typically rely on the effects of projective measurements on single quantum systems, and the absence of true projective measurements in ensemble NMR systems makes this essentially impossible. More advanced quantum communication protocols, such as E91 quantum cryptography \cite{Ekert1991} and quantum teleportation \cite{Bennett1993} rely on distributing entanglement over significant distances \cite{Nadlinger2022,Zhang2022b}. This is not really possible in NMR, where the entanglement is confined within a single molecule, and although the teleportation circuit has been demonstrated in a three spin system \cite{Nielsen1998}, the information was only moved over a few angstroms.

The situation for quantum sensing with NMR is the reverse: here significant results have been demonstrated for entanglement-enhanced magnetic field sensing \cite{Jones2009, Simmons2010, Shukla2014}, but these experiments are in reality little more than relabelled versions of the traditional HMQC \cite{Mueller1979} and HSQC \cite{Bodenhausen1980} experiments, reflecting the close relationship between Schr\"odinger Cat states and maximal multiple quantum coherences \cite{Knill2000}.

\subsection{Structure and scope}
In my first review in this journal \cite{Jones2001a} I provided a general introduction to quantum computation and the main methods used for implementing it in NMR spin systems, while my second review \cite{Jones2011} sought to provide a fairly complete summary of all the major experimental approaches in use at that time. These two reviews bracket a very busy period in which rapid progress was made and a large number of papers were published by many different research groups. Since 2011 the field has become quieter, with many of the remaining researchers tending to concentrate on a small number of particular topics. In this review I will begin with a brief introduction, followed by a summary of popular spin systems, and will then concentrate on some areas of current interest. These mostly relate to \textit{quantum control}, that is the design of composite pulses, shaped pulses, and pulse sequences, to perform particular transformations of quantum states \cite{Vandersypen2005}.

Throughout the text I will assume familiarity with conventional NMR methods and with elementary quantum mechanics, but no detailed familiarity with quantum information theory. I will, however, discuss some conventional NMR themes in the context of quantum information, in part to clarify how the two notations interrelate, but also to indicate some limitations on the situations in which these conventional NMR techniques can be applied.

\section{DiVincenzo criteria}
The suitability of any physical system for building a quantum computer is traditionally assessed using the five DiVincenzo criteria \cite{DiVincenzo2000}, briefly summarised in Table~\ref{tab:DVC}. Although this list is arguably not the best way to think about realistic proposals \cite{Ladd2010}, it does provide a simple structure enabling different physical technologies to be easily compared. As we will see for NMR, the central conclusion is that while the construction of small demonstration systems is straightforward, there are enormous difficulties in scaling these up to the sizes required for a genuinely useful device.

\begin{table}
\begin{center}
\begin{tabular}{|l|l|}
\hline
Criterion&NMR implementation\\\hline
1. qubits&spin-\half\ nuclei\\
2. initialisation&pseudo-pure states\\
3. low decoherence&long $T_2$\\
4. logic gates&pulses and delays\\
5. measurement&NMR spectrum\\
\hline
\end{tabular}
\end{center}
\caption{A summary of the 5 DiVincenzo criteria and how they might be met in NMR systems. Although all 5 criteria are met well enough for simple demonstrations, none of them are met in a genuinely scalable way.}\label{tab:DVC}
\end{table}

\paragraph{1. A scalable physical system with well characterized qubits} The basic approach in NMR is simple, using a single spin-\half\ nucleus in a small molecule to represent each qubit. I will mostly not consider proposals which use electron spins \cite{Platzman1999, Morton2005a, Byeon2021} or which combine electron and nuclear spin qubits \cite{Kane1998, Rahimi2005, Sato2007}. I will also not consider proposals involving high-spin nuclei, such as schemes that represent a qutrit using a spin-1 nucleus in a liquid crystal solvent \cite{Das2003c,Dogra2016} or schemes that use the four levels of a spin-$\frac32$ nucleus \cite{Khitrin2000,Sinha2001,Ermakov2002} or the eight levels in a spin-$\frac72$ nucleus \cite{Khitrin2001, Murali2002} to represent two or three qubits in one system. Similarly, I will largely only consider small molecules in isotropic liquids, rather than systems in the solid state \cite{Cory2000, Leskowitz2003, Baugh2005, Baugh2006, Ryan2008a} or systems with partial local ordering induced by liquid crystal solvents \cite{Yannoni1999, Marjanska2000, Fung2001, Fung2001a, Mahesh2002, Das2004b, Lee2004, Lee2005, Lee2005a, Mahesh2006, Lee2007, Lu2010, Li2014}.

As discussed in Section~\ref{sec:spinsystem}, it is straightforward to find suitable spin systems to represent small numbers of qubits, but the difficulty increases sharply with the number of spins required. This is the first reason why conventional NMR does not provide a realistic route to a useful quantum computer.

\paragraph{2. The ability to initialize the state of the qubits to a simple fiducial state, such as \ket{000\dots}} In many approaches to quantum computing this is done by some sort of cooling process: sometimes by direct cooling to the energetic ground state, but more frequently by indirect approaches, such as optical pumping, which allow a chosen state to be selectively prepared \cite{Schindler2013}. Cooling is a generally impractical approach for NMR quantum computing, not because the temperatures required (of the order of mK) are unattainable, but rather because the sample must normally be kept in the liquid state to obtain the desired motionally averaged Hamiltonian. While a wide range of signal enhancement approaches have been demonstrated \cite{Eills2023}, which reduce the effective spin temperature while keeping the molecular lattice close to room temperature, the enhancements obtainable are not normally high enough to reach the desired pure spin states \cite{Jones2000a}. The sole exception to this is the use of \textit{para}-hydrogen \cite{Hubler2000}, but as yet this has only been used to produce pure states for two-spin systems \cite{Anwar2004}.

Instead of preparing pure states the standard approach for NMR quantum computing is to prepare \textit{pseudo-pure states}, also called \textit{effective pure states} \cite{Cory1996, Cory1997, Cory1998a, Gershenfeld1997, Chuang1998b}, as discussed in Section~\ref{sec:PPS}. This process cannot be performed scalably \cite{Warren1997, Gershenfeld1997a}, once again limiting NMR QIP to relatively small spin systems.

\paragraph{3. Long relevant decoherence times, much longer than the gate operation time} In NMR implementations this means that the slowest interactions used to implement gates, usually the scalar couplings between spins, must be fast compared with the fastest relaxation time, usually taken as the spin--spin relaxation time, $T_2$, although in reality the relaxation times of multiple quantum coherences may be more relevant. Naively this means that coupling patterns must be well resolved, but this is a sufficient rather than a strictly necessary condition, as inhomogeneous broadening, which makes $T_2^*$ less than $T_2$, can be refocused \cite{Hahn1951}. 

However it is important to realise that \textit{much longer} in this requirement means above the fault tolerant threshold \cite{Knill1998b}. This threshold depends on the error correction code chosen and the overhead one is prepared to tolerate \cite{Fowler2012}, but in practice a ratio of at least 100 is essential and a factor closer to 10,000 is preferable. Even the lower limit is challenging, and the higher ratio is far out of reach, and so performing extended quantum computations with NMR is not currently possible.

\paragraph{4. A ``universal'' set of quantum gates} Gate universality, which is the ability to approximate any desired evolution using a network of gates from some finite set, is a much studied topic in QIP. Very early papers assumed that three-qubit gates would be required \cite{Deutsch1989}, but a key early result was that two-qubit gates suffice \cite{DiVincenzo1995, Barenco1995a}, and indeed that almost any two-qubit gate is universal \cite{Deutsch1995, Lloyd1995}. More practically the combination of a universal set of single-qubit gates and any non-trivial two-qubit gate, such as the \CNOT\ gate \cite{Barenco1995b}, is universal \cite{Barenco1995}. It can also be shown that two particular gates, traditionally taken as the Hadamard gate and the fourth root of Z gate, suffice to form a universal set of single-qubit gates \cite{Boykin2000}. More importantly for NMR implementations, the set of single-spin rotations around axes in the $xy$-plane, corresponding to the set of spin-selective pulses, combined with free evolution in the presence of scalar coupling interactions, is universal \cite{Jones1998b}.

As hinted at above, one central problem for gate implementation in NMR QIP is the problem of spin-selective excitation. Most other proposals for implementing quantum computation ultimately rely on some form of spatial selection, in which different qubits are implemented using physical systems in different regions of space, but this is not possible in NMR systems, which are built around a macroscopic ensemble of rapidly tumbling systems. Instead the qubits are distinguished using their different resonance frequencies.

Such frequency selection is trivial in heteronuclear spin systems, but there are only a finite number of spin-\half\ nuclei available. In homonuclear spin systems the chemical shift interaction provides sufficient dispersion to distinguish small numbers of qubits, but the finite range of chemical shifts once again limits this approach to a fairly small number of spins of any one nuclear species \cite{Jones2000a}. Some common homonuclear and heteronuclear spin systems are discussed in Section~\ref{sec:spinsystem}, and the design of robust spin-selective rotations is a central feature of Sections \ref{sec:control}, \ref{sec:OC} and \ref{sec:GRAPE}.

A second central problem is the design of refocusing networks to remove unwanted spin--spin couplings. Although free evolution under the natural background Hamiltonian is formally universal when combined with single-qubit gates, it does not normally correspond naturally to a conventional logic gate. More fundamentally NMR quantum computers differ from most other designs in that these logic gates are ``always on'', and have to be turned off when they are not required \cite{Bhole2020a}. Approaches for doing this efficiently are discussed in Section~\ref{sec:WHSE}. Related to this is the problem of turning off couplings to spins outside the spin system used for information processing. In conventional NMR this is usually achieved by decoupling, but within QIP it can be more appropriate to use \textit{dynamical decoupling}, in which the refocusing pulses are applied to the system (the spins of interest) rather than the surroundings (their coupling partners), as explored in Section~\ref{sec:dd}.

\paragraph{5. A qubit-specific measurement capability} Qubit measurement is obviously important as there is no point in performing a computation if the result cannot be read out in some way. However quantum measurement is very different from classical measurement. In the classical world a measurement can be thought of as revealing a pre-existing state of a classical object, and can be performed without affecting the state, but quantum measurement is nothing like this \cite{NCbook, Merminbook}. Every measurement process has an associated set of outcomes, which form a complete orthonormal basis for the system, and the result of a measurement is to project the system at random into one of these possible outcome states, with the outcome probabilities given by the square moduli of the corresponding amplitudes. For a measurement performed in the computational basis only these basis states can be measured non-intrusively: any measurement on a superposition state will return one of the contributing basis states at random, with any entanglement in the superposition reflected in correlations between different bits in the outcome.

In NMR quantum computing, measurement is achieved by observing the NMR spectrum, either directly or after applying excitation pulses to one or more spins. This is not a true quantum measurement, but rather the determination of an ensemble averaged expectation value for some traceless observable \cite{Cory1997}. If the spin system is in an eigenstate before the measurement then this state can be identified from the intensities of lines in appropriate multiplets \cite{Jones2011}, and in some special cases the ensemble nature of NMR can be useful \cite{Jones1999a}. However, for algorithms which result in a superposition of possible answers, one of which is selected at random by the measurement process, ensemble averaged results are not useful, and in NMR implementations of such algorithms it is common to note simply that the observed NMR signal matches the simulated predictions \cite{Vandersypen2001}. For quantum protocols that result in entangled states \cite{Cummins2000a}, which can be related to multiple quantum coherences \cite{Jones2001a, Jones1998b}, the outcome may be particularly difficult to monitor directly, although in some cases useful simple measurements can be found \cite{Anwar2004, Bhole2020}.

One way to overcome this is to use quantum state tomography, in essence measuring enough different observables that it is possible to completely reconstruct the density matrix, or at least its traceless part, the deviation density matrix \cite{Chuang1998b, Laflamme1998, Vandersypen1999, Teklemariam2001, Boulant2002}. Several methods have been used to increase the efficiency of quantum state tomography in NMR \cite{Long2001, Das2003, Das2003a}, and more generally \cite{Linden2002, Linden2002a, Xin2017}, but the exponential growth in the number of elements in the full density matrix makes complete reconstructions very challenging for large spin systems.

Furthermore, the lack of projective measurements means that qubits cannot be easily reset. Quantum error correction protocols \cite{Shor1995, Steane1996, Acharya2023} depend on access to \textit{ancilla} qubits in a well-defined state, typically \ket{0}, to record the errors which have occurred. The error correction process needs to be carried out repeatedly, which requires either that the ancillas are reset to their initial state or a continuous supply of fresh ancillas is available. Although single rounds of error correction have been demonstrated in NMR \cite{Cory1998, Knill2001}, the absence of a reset process renders effective error correction difficult in NMR systems \cite{Steane2004}.

\section{States} \label{sec:states}
There is an exact correspondence between the pure states of an isolated spin-\half\ nucleus and a qubit, and both are commonly described using the Bloch sphere picture. For a single qubit a general state can be written as
\begin{equation}
\ket\psi=c_0\ket0+c_1\ket1\label{eq:alphabeta}
\end{equation}
where $c_0$ and $c_1$ are complex numbers, subject to the normalisation constraint that
\begin{equation}
|c_0|^2+|c_1|^2=1.
\end{equation}
Given an ensemble of identical copies of this system experiments can be performed which provide information on the magnitudes of $c_0$ and $c_1$, and on their \textit{relative} phase, but there is no method whatsoever to obtain any information on the \textit{absolute} phases of $c_0$ and $c_1$. Equivalently, the state \ket\psi\ is completely indistinguishable from the state
\begin{equation}
\ket{\psi'}=\rme^{\rmi\gamma}\ket\psi=\rme^{\rmi\gamma}c_0\ket0+\rme^{\rmi\gamma}c_1\ket1,
\end{equation}
so the \textit{global phase} $\gamma$ has no physical meaning. One common approach is to choose $\gamma$ so that the amplitude of the \ket0 component is real and positive, which combined with normalisation enables a single qubit to be described as
\begin{equation}
\ket\psi=\cos(\theta/2)\ket0+\rme^{\rmi\phi}\sin(\theta/2)\ket1,
\end{equation}
with $0\le\theta\le\pi$ and $0\le\phi<2\pi$. Thus any state of a single qubit can be represented using spherical polar coordinates as a point on the surface of a unit sphere, which is the Bloch sphere.
\begin{figure}
\begin{center}
\includegraphics{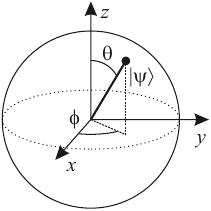}
\end{center}
\caption{Representing a pure state of a single qubit as a point on the surface of the Bloch sphere using spherical polar coordinates. This is entirely equivalent to the Bloch sphere used in conventional NMR , where Cartesian coordinates are more common.} \label{fig:BlochSphere}
\end{figure}

Exactly the same approach can be used within NMR, with the eigenstates $\ket\alpha=\ket{+\half}$ and $\ket\beta=\ket{-\half}$ of a spin-\half\ nucleus playing the roles of \ket0 and \ket1, and the Bloch vector simply connecting the origin and an appropriate point on the Bloch sphere. The main difference is that the states used in NMR are mixed states, and so strictly lie within the Bloch sphere rather than on its surface. It is, however, common to ignore this, as discussed below. The use of NMR operator notation also leads to the Bloch vector normally being described in Cartesian notation rather than spherical polars.

\subsection{Mixed states}
The states described in equation~\ref{eq:alphabeta} are \textit{pure states}, which correspond to the quantum system being in a single well defined state. This state need not be an eigenstate, but it is unitarily equivalent to an eigenstate, as there will always be some unitary transformation that interconverts \ket0 and \ket\psi. A more general possibility is that the qubit can be in a \textit{mixed state}, which is not a single well defined state but rather a probabilistic mixture of such states.

Mixed states cannot be described using kets, but are instead described using density matrices of the form
\begin{equation}
\rho=\sum_j p_j \ket{\psi_j}\bra{\psi_j},
\end{equation}
where the $p_j$ are probabilities, and so must be real numbers with $0\le p_j\le1$ and $\sum_j p_j=1$. This form shows that density matrices must be Hermitian (that is $\rho=\rho^\dagger$), and so must have an orthonormal eigenbasis \cite{NCbook}. They must also be positive semidefinite, which means that their eigenvalues must be non-negative, that is positive or zero. Two important special cases are pure states, which have a single eigenvalue equal to $1$ with the rest being $0$, and the \textit{maximally mixed state}, which is an equal mixture of all the eigenstates of the system. For a qubit this takes the form
\begin{equation}
\half E=\begin{pmatrix}\half&0\\0&\half\end{pmatrix}.
\end{equation}

For a single qubit the situation is particularly simple. Considering the state in its eigenbasis it is clear that any mixed state can be written in the form
\begin{equation}
\rho=p\ket{\psi}\bra{\psi}+(1-p)\ket{\psi^\perp}\bra{\psi^\perp},\label{eq:rhoqubita}
\end{equation}
for some state \ket\psi, where
\begin{equation}
\ket{\psi^\perp}=c_1^*\ket0-c_0^*\ket1
\end{equation}
is the state orthogonal to \ket\psi, and we can choose the states such that \ket\psi\ has a probability equal to or greater than that of \ket{\psi^\perp}, so that $\half\le p\le1$, with a pure state corresponding to $p=1$. In particular the maximally mixed state can be decomposed not just as an equal mixture of \ket0 and \ket1, but also as an equal mixture of any state and its orthogonal partner,
\begin{equation}
\half E=\half\ket\psi\bra\psi+\half\ket{\psi^\perp}\bra{\psi^\perp}.
\end{equation}
This allows equation~\ref{eq:rhoqubita} to be rewritten as
\begin{equation}
\rho=2(1-p)\,\half E+(2p-1)\ket{\psi}\bra{\psi},\label{eq:rhoqubitb}
\end{equation}
corresponding to a mixture of the maximally mixed state and an excess population of \ket\psi. This means that every state of a single qubit is a pseudo-pure state. Since the maximally mixed state gives no signal in NMR experiments the behaviour of $\rho$ is almost indistinguishable from that of the corresponding pure state \ket\psi, differing only in a reduced signal intensity. For this reason it is common within NMR to treat mixed states of single spins as if they were pure states. However it is necessary to be much more careful when describing systems with multiple spins, as discussed in  Section~\ref{sec:PPS}.

Within conventional NMR a different but related description is normally used. The excess component can be rewritten using
\begin{equation}
\begin{split}
\ket{\psi}\bra{\psi}=&
\half\left(\ket{\psi}\bra{\psi}+\ket{\psi^\perp}\bra{\psi^\perp}\right)\\
+&\half\left(\ket{\psi}\bra{\psi}-\ket{\psi^\perp}\bra{\psi^\perp}\right)\\
=&\half E+I_\psi
\end{split}
\end{equation}
where $I_\psi$ is an angular momentum operator parallel to \ket\psi, defined by
\begin{equation}
I_\psi=\sin\theta\cos\phi\,I_x+\sin\theta\sin\phi\,I_y+\cos\theta\,I_z,
\end{equation}
with Cartesian components corresponding to the Bloch vector. Thus
\begin{equation}
\rho=\half E+(2p-1)I_\psi,\label{eq:rhoqubitc}
\end{equation}
where the fact that $I_\psi$ is traceless ensures that the maximally mixed term is always $\half E$ to get the correct trace. The conventional NMR approach is then to drop not only the maximally mixed state but also the term describing the size of the polarisation, here written as $2p-1$, or equivalently to assume that $p=1$, and so describe the spin state as $I_\psi$. While this simplified approach can be highly successful it must be remembered that angular momentum operators are not proper density matrices, as they are not positive semidefinite with unit trace, and so cannot always be naively substituted into formulae derived for density matrices.

\subsection{Fidelities}\label{sec:fidelities}
The concept of \textit{state fidelity} is an important one in quantum information theory, providing a measure of how similar two quantum states are. For two pure states it is defined simply as the square modulus of the inner product,
\begin{equation}
F_{\psi,\phi}=|\inner{\psi}{\phi}|^2=\inner\psi\phi\inner\phi\psi,\label{eq:Fpure}
\end{equation}
which has limiting values $F=1$ when $\ket\phi=\ket\psi$ and $F=0$ when $\ket\phi=\ket{\psi^\perp}$. This definition extends by linearity to give the fidelity between a pure state and a mixed state,
\begin{equation}
F_{\psi,\rho}=\bra\psi\rho\ket\psi.\label{eq:Fmix}
\end{equation}
The extension to comparing two mixed states, $\rho$ and $\sigma$, is more complicated, and the naive generalisation $\textrm{Tr}(\rho\sigma)$ is not suitable. The correct fidelity in this case is the Uhlmann--Jozsa fidelity \cite{Uhlmann1976, Jozsa1994} which is defined as
\begin{equation}
F_{\rho,\sigma}=\left[\textrm{Tr}\left(|\sqrt{\rho}\sqrt{\sigma}|\right)\right]^2, \label{eq:FUJ}
\end{equation}
where the modulus of an operator is defined by
\begin{equation}
|A|=\sqrt{AA^\dag}.
\end{equation}
Note that all proper density matrices are Hermitian and positive semidefinite, and so $\sqrt\rho$ and $\sqrt\sigma$ exist, and are also Hermitian and positive semidefinite. This leads to the more usual form
\begin{equation}
F_{\rho,\sigma}=\left[\textrm{Tr}\left(\sqrt{\sqrt{\rho}\,\sigma\sqrt{\rho}}\right)\right]^2.\label{eq:Fclassic}
\end{equation}

The fearsome appearance of this equation, especially to readers who are unaccustomed to matrix square roots, has led to many attempts to find simpler formulae \cite{Mendonca2008, Liang2019}, but none of these fulfil all of the six properties achieved by the Uhlmann--Jozsa fidelity \cite{Jozsa1994}, four of which are essential and two of which are highly desirable. In particular a fidelity should lie between 0 and 1, achieving a value of 1 if and only if $\rho=\sigma$, should be symmetric between $\rho$ and $\sigma$, should be invariant under unitary transformations, and should reduce to the form of equation~\ref{eq:Fmix} when $\rho$ or $\sigma$ is pure.

\begin{figure}
\begin{center}
\includegraphics{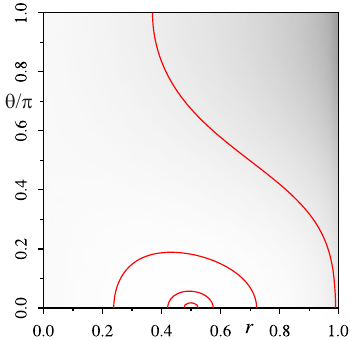}
\end{center}
\caption{The Uhlmann--Jozsa fidelity between the target state, equation~\ref{eq:rho34}, and the general state, equation~\ref{eq:sigma} for the case $\phi=0$, plotted over the range $0\le r\le1$, and $0\le\theta\le\pi$. Contours are plotted at fidelities of 0.9, 0.99, 0.999, and 0.9999, revealing a clear maximum at $r=0.5$, $\theta=0$, corresponding to $\rho=\half E+\half I_z$.}\label{fig:fUJ}
\end{figure}

The naive generalisation $\textrm{Tr}(\rho\sigma)$ does not meet these requirements: consider the simple example
\begin{equation}
\rho=\begin{pmatrix}{\textstyle\frac{3}{4}}&0\\0&{\textstyle\frac{1}{4}}\end{pmatrix}\label{eq:rho34}
\end{equation}
for which $\textrm{Tr}(\rho^2)=\frac{5}{8}$, showing that this form does not reach a value of 1 for $\rho=\sigma$. The highest value which can be reached by any proper density matrix is achieved by
\begin{equation}
\sigma=\begin{pmatrix}1&0\\0&0\end{pmatrix}
\end{equation}
for which  $\textrm{Tr}(\rho\sigma)=\frac{3}{4}$. It is also impossible to ``patch up'' this definition without introducing other problems. In contrast the Uhlmann--Jozsa fidelity behaves correctly. This can be seen by calculating the fidelity between equation~\ref{eq:rho34} and a general mixed state written in NMR notation
\begin{equation}
\sigma=\half E+r\left(\sin\theta\cos\phi I_x+\sin\theta\sin\phi I_y+\cos\theta I_z\right)\label{eq:sigma}
\end{equation}
for which the fidelity is easily seen to be independent of $\phi$, so without loss of generality we can assume $\phi=0$. Plotting this fidelity as a function of $r$ and $\theta$, as shown in Figure~\ref{fig:fUJ}, gives a clear maximum at $r=\half$ and $\theta=0$, where a level of 1 is achieved, exactly as expected.

Thus it appears that the square roots cannot be entirely avoided, but it is possible to recast the Uhlmann--Jozsa fidelity into a different form which is much easier to calculate numerically. In particular it can be shown \cite{Baldwin2023} that provided $\rho$ and $\sigma$ are proper density matrices then the form
\begin{equation}
F_{\rho,\sigma}=\left[\textrm{Tr}\left(\sqrt{\rho\sigma}\right)\right]^2\label{eq:Fnew}
\end{equation}
can be used instead. Furthermore it is not actually necessary to explicitly find $\sqrt{\rho\sigma}$ as only its trace, which is equal to the sum of its eigenvalues, is required, and it can be shown that these eigenvalues are equal to the square roots of the eigenvalues of $\rho\sigma$. Using this efficient approach is it possible to speed up the computation of the Uhlmann--Jozsa fidelity by around a factor of ten \cite{Baldwin2023}.

\section{Choice of spin system}\label{sec:spinsystem}
When choosing a spin system for implementing an NMR quantum computation it is necessary to find a molecular system containing the right number of spin-\half\ nuclei in a coupled network. It is not necessary that all the nuclei be directly coupled, but it is necessary that they all be connected directly or indirectly \cite{Collins2000} by some chain of sufficiently large couplings.

The conceptually simplest approach is to use an entirely heteronuclear spin system, as this makes selective addressing trivial, but this is limited by the small number of suitable spin-\half\ nuclei, and so many implementations are at least partly homonuclear, containing two or more spins of a particular nuclear species. With homonuclear systems a key decision is whether to work with all the spins of a given type in the same rotating frame, or to assign a separate frame to every spin, sometimes called abstract reference frames \cite{Knill2000}. This decision can be sidestepped when there are only two spins of any given type, as in this case the two abstract frames will align at stroboscopic intervals \cite{Jones1998c}. In principle the same decision must be made for fully heteronuclear systems, but here the universal practice is to assign each nuclear species its own rotating frame, usually at or close to resonance with the single spin of that type.

A further consideration in homonuclear systems is whether the spin--spin couplings can be treated as weak. In practice this point is frequently ignored and a weak-coupling Hamiltonian is regularly assumed even when deviations are clearly visible in the NMR spectrum. This is not, of course, a concern in heteronuclear systems.

\subsection{Choosing nuclei}
While there are a large number of spin-\half\ nuclei which could in principle be used, the choice in practice is strongly influenced by easy availability of certain chemical systems \cite{Glaser2001} and of commercial NMR equipment. There are six spin-\half\ nuclei which occur with near 100\% natural abundance, but of these only three (\nuc{1}{H}, \nuc{19}{F}, and \nuc{31}{P}) have the chemical versatility to  be easily included in small organic molecules, with the other three (\nuc{89}{Y}, \nuc{103}{Rh}, and \nuc{169}{Tm}) being metals. To this short list can be added \nuc{13}{C} and \nuc{15}{N}, reflecting the relatively easy availability of selective isotopic labelling and the wide availability of suitable double, triple and quadruple resonance probes for chemical and biochemical studies. In various combinations these five nuclei completely dominate spin-\half\ quantum computing experiments. In one extreme case a fully heteronuclear five-qubit computer was designed using all five nuclei \cite{Marx2015, Silva2016}, which required the use of a custom six-channel probe (including the \nuc{2}{H} lock channel) \cite{Marx2015}.

Use of other spin-\half\ nuclei has been far more limited. A wide range of exotic spins have been discussed from a theoretical perspective but without experimental demonstrations \cite{Lino2020,Lino2022}. The most important experimental example is \nuc{29}{Si}, which has been used in star-topology systems, in which a single \nuc{29}{Si} nucleus is surrounded by 12 \cite{Simmons2010} or even 36 \cite{Pande2017,Pal2018} \nuc{1}{H} nuclei. By making all NMR measurements at the \nuc{29}{Si} frequency the experiment is only sensitive to the 5\% of the sample containing a \nuc{29}{Si} nucleus, thus automatically selecting a labelled subset of molecules.

This trick cannot be easily extended to systems containing two or more such nuclei, limiting its applicability. A system containing two silicon atoms will appear in the \nuc{29}{Si} spectrum as an equal mixture of the two different ``singly labelled'' isotopomers, with much weaker signals from the rare doubly labelled compound. As each isotopomer gives rise to its own multiplet it is simple to separate the two signals, permitting easy study of either of the two spin systems. This approach is quite widely used with natural abundance \nuc{13}{C} to extend a spin system comprising \nuc{1}{H} or \nuc{19}{F} nuclei in an organic molecule, in effect adding a single \nuc{13}{C} nucleus without explicit labelling.

As well as considering the spin-system used to represent quantum information it is also necessary to ensure that any other spins in the molecule can be ignored. Clearly spin-0 nuclei, such as \nuc{16}{O}, can be entirely ignored, and high spin nuclei, such as \nuc{2}{H}, \nuc{14}{N}, and \nuc{35/37}{Cl}, can be largely ignored, as their rapid quadrupolar relaxation acts to remove the effects of any couplings to the spin-\half\ nuclei of interest. Furthermore, labile \nuc{1}{H} nuclei can be easily exchanged for \nuc{2}{H} by dissolving in $\textrm{D}_2\textrm{O}$.

It is also possible to ignore spin-\half\ nuclei which are not coupled to the main spin system: although such spins are visible in NMR spectra they will not affect the evolution of the spins of interest. Here ``not coupled'' really means having a coupling constant low enough to ignore, which is a practical question rather than a matter of principle. For example, the fully heteronuclear five-qubit computer mentioned above also contains two N-methyl and two O-ethyl groups which are weakly coupled to the main system. Most of these couplings are under 1\,Hz, and can be ignored, but the largest long range couplings were decoupled using selective pulses \cite{Marx2015}.

\subsection{Systems with two spins}
\begin{figure}
\begin{center}
\includegraphics{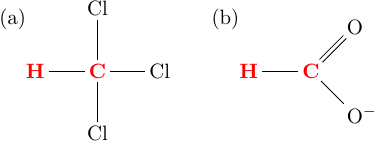}
\end{center}
\caption{Popular two qubit heteronuclear systems include (a) chloroform and (b) the formate anion, both with \nuc{13}{C} labelling. Nuclei used as qubits are shown in red boldface, and the other nuclei can be neglected.} \label{fig:2qubit}
\end{figure}

A two-spin system can only be either homonuclear or fully heteronuclear, and both approaches have proved popular. The first NMR quantum computing experiments were performed using either a pair of \nuc{1}{H} nuclei in cytosine dissolved in $\textrm{D}_2\textrm{O}$ \cite{Jones1998c, Jones1998d, Jones1999a} or the combination of a \nuc{1}{H} and a \nuc{13}{C} nucleus in \nuc{13}{C} labelled chloroform dissolved in  acetone-$\textrm{d}_6$ \cite{Chuang1998, Chuang1998a, Childs2001} or $\textrm{CDCl}_3$ \cite{Jones2000b}, see Figure~\ref{fig:2qubit}.

Many other HH systems have been studied, including 2,3-dibromothiophene \cite{Cory1998a}, uracil \cite{Linden1999b}, 5-nitrofuraldehyde \cite{Dorai2000a,Arvind2001,Mahesh2001}, coumarin \cite{Dorai2000a}, and 5-bromothiophene-2-carbaldehyde \cite{Roy2010}, as well as a range of systems synthesised from \textit{para}-hydrogen \cite{Hubler2000, Anwar2004, Anwar2004b, Anwar2004a}. Systems involving a pair of coupled \nuc{31}{P} nuclei have also been explored \cite{Ito2009}.

For heteronuclear systems the choice of combining \nuc{1}{H} with \nuc{13}{C} is very obvious, but the early choice of chloroform has some disadvantages related to the relaxation of the \nuc{13}{C} nucleus. This has a shortened $T_2$, arising from scalar relaxation of the second kind \cite{Abragam1961} caused by rapid quadrupolar relaxation of directly bonded \nuc{35/37}{Cl} nuclei, which limits the number of quantum gates that can be performed. This is combined with a very long $T_1$, limiting the repetition rate if experiments are started from the thermal equilibrium state. A popular alternative HC system with slightly more balanced relaxation times is provided by labelled sodium formate in $\textrm{D}_2\textrm{O}$ \cite{Leung1999,Xiao2005,Xiao2006a,Fitzsimons2015,Greganti2021}, or the closely related formic acid \cite{Jiang2018a}, see Figure~\ref{fig:2qubit}. Experiments have also been demonstrated with labelled dimethylformamide \cite{Das2003a}, where the methyl protons can simply be ignored. However, chloroform remains the overwhelmingly popular choice \cite{Peng2001, Zhu2001a, Das2002, Peng2003, Peng2003a, Peng2005, Peng2005a, Du2006b, Zhang2007, Peng2007a, Souza2008, Du2010, Athalye2011, Chen2011a, Katiyar2012, Lu2012a, Roy2012, Hegde2014, Maciel2015, Luo2016, Li2016, Wang2016, Xin2017a, Wang2018, Wang2018a, Bian2019, Singh2019a, Chen2020, Xin2020b, Yang2020b, Yang2021, Zhao2021, Liu2021, Li2022, Lin2022}.

Other heteronuclear combinations are less popular, perhaps just because suitable probes are not quite so widely available. The combination of \nuc{1}{H} and \nuc{19}{F} has been demonstrated in 5-fluorouracil \cite{Dorai2001}, a convenient and readily available heteronuclear replacement for uracil. Perhaps more interesting is the combination of \nuc{1}{H} and \nuc{31}{P}, which was originally demonstrated in phosphonic acid \cite{Fu1999,Long2001a}, which has a particularly large scalar coupling (almost 650\,Hz) between \nuc{31}{P} and the directly bonded \nuc{1}{H}. This system has subsequently been adapted to build a tabletop two-qubit NMR device, called SpinQ Gemini \cite{Hou2021, Varga2023}, based around dimethylphosphite, where the one bond coupling of almost 700\,Hz dominates over the long-range couplings to the methyl protons. An even larger coupling, over 850\,Hz, is found between the directly bonded \nuc{19}{F} and \nuc{31}{P} nuclei in sodium fluorophosphate \cite{Krithika2019, Pal2020, Krithika2021, Sharmila2022}.

\subsection{Systems with three spins}
\begin{figure}
\begin{center}
\includegraphics{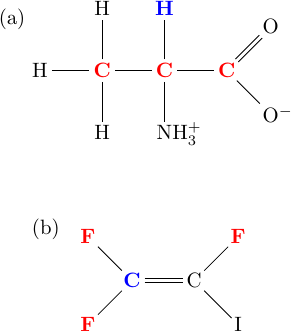}
\end{center}
\caption{Popular three qubit homonuclear systems include (a) \nuc{13}{C} labelled alanine and (b) iodotrifluoroethylene. The three main qubits are shown in red boldface, but these molecules have also been extended to four qubit partly heteronuclear systems by including the nuclei shown in blue boldface.} \label{fig:3qubithom}
\end{figure}
A wide range of different three-spin systems have been explored. Fully homonuclear systems (Figure~\ref{fig:3qubithom}) have been led by studies of the three \nuc{13}{C} spins in labelled alanine \cite{Teklemariam2001, Cory1998, Collins2000, Tseng1999, Nelson2000, Kim2000, Viola2001, Weinstein2001, Peng2002, Teklemariam2002, Teklemariam2003, Fortunato2002, Weinstein2002, Kim2002, Xiao2002, Lee2002a, Xue2002, Weinstein2004, Du2006a, Fitzsimons2007, Kondo2007, Ren2009, Zhu2011, Bagnasco2014, Wei2014a}, but three \nuc{1}{H} spins in 2,3-dibromopropanoic acid \cite{Linden1998, Linden1999a,Mahesh2001a} or in chlorostyrene \cite{Du2000a, Du2001a} or three \nuc{19}{F} spins in bromotrifluoroethylene \cite{Vandersypen1999}, 2,3,4-trifluoroaniline \cite{Mangold2004}, 4-bromo-1,1,2-trifluoro-1-butene \cite{ Mitra2008}, or iodotrifluoroethylene \cite{Du2007, Peng2008b, Golze2012, Katiyar2013, Shukla2013, Rao2013, Joshi2014, Shukla2014a, Hegde2014a, Dogra2015, Katiyar2016, Li2016a, Peterson2016, Devra2018, Pal2019a, Feng2022, Singh2022b} have also proved popular. Although \nuc{19}{F} probes are less widely available than \nuc{1}{H}, the wide range of chemical shifts and the large size of the scalar couplings makes \nuc{19}{F} a tempting choice \cite{Vandersypen1999}.

\begin{figure}
\begin{center}
\includegraphics{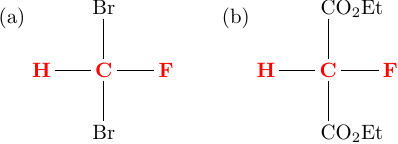}
\end{center}
\caption{Popular three qubit heteronuclear systems include (a) di\-bromo\-fluoro\-methane and (b) diethyl-fluoromalonate, here drawn to emphasise the similarity of the two systems. Note that the protons in the ethyl groups are not significantly coupled to the main qubits and give signals well separated from the \nuc{1}{H} qubit multiplet.} \label{fig:3qubithet}
\end{figure}
Among fully heteronuclear implementations (Figure~\ref{fig:3qubithet}) the most popular approach is to combine \nuc{1}{H}, \nuc{13}{C}, and \nuc{19}{F} nuclei in \nuc{13}{C} labelled di\-bromo\-fluoro\-methane \cite{Vandersypen2000, Mitra2007, Samal2011, Bhole2016, Pal2019, Pal2021, Krithika2021, Sharmila2022}, ethyl 2-fluoroacetoacetate \cite{Zhang2007a, Kampermann2010}, or diethyl-fluoromalonate \cite{Zhang2008, Peng2008, Peng2009, Peng2010, Peng2010a, Chen2011, Lu2011, Li2011, Wu2012, Lu2012, Feng2013, Wu2013a, Zheng2013, Gao2013, Hou2014a, Xin2015, Jin2016, Ma2016, Ma2016a, Li2017d, Xin2017b, Zheng2018, Singh2018, Singh2019, Ji2019, Zhu2019, Gautam2020, Yang2020a, Singh2020, Ding2021, Singh2022, Singh2022a, Singh2023}. Although some studies of diethyl-fluoromalonate explicitly refer to \nuc{13}{C} labelling \cite{Li2011, Singh2019} it appears that some other experiments were performed with unlabelled samples, although it is only rarely that this is clearly described \cite{Zhang2008}.

Between the extremes of homonuclear and fully heteronuclear systems lie the mixed systems, with two spins of one nuclear type and the third of another. This approach allows the unique spin to be directly controlled while stroboscopic methods can be applied to the two spins of the same species, and can allow a convenient distinction between different roles for particular spins, for example for input and output. A HHF system has been explored in 4-fluoro-7-nitro-benzofuran \cite{Dorai2000a,Das2004}, while HHP has been studied using \textit{E}-(2-chloroethenyl)phosphonic acid \cite{Cummins2002} and HHN using \nuc{15}{N} labelled acetamide \cite{Khaneja2007,Wei2014}. Among doubly labelled compounds the most popular approach has been to use the HCC system, usually in trichlororethene \cite{Nielsen1998, Cory1998, Laflamme1998, Miquel2002, Zhang2005, Zhang2006, Liu2008a, Lu2014, Atia2014, Brassard2014, Atia2016, Lu2016, Li2017b, Li2017c} but sometimes in tris(trimethylsilyl)silane-acetylene \cite{Henry2006, Henry2007, Ryan2009, Park2012} or in propyne \cite{Teklemariam2003a}.

\subsection{Systems with four spins}
With four spins the range of possibilities becomes very large, and here I list only some notable examples. An early experiment used 1-chloro-2-nitrobenzene as an HHHH system \cite{Cory1998a}, but only used this to control three qubits to demonstrate a Toffoli gate. Similar results were shown using 2,3-difluoro-6-nitrophenol as an HHFF system \cite{Mahesh2001,Das2004,Das2004a} and \nuc{13}{C} labelled alanine as an HCCC system \cite{Du2001a, Peng2004}, with selective decoupling of the methyl protons to simplify the spin system. More sophisticated experiments were performed using glycine as an HNCC spin system \cite{Ollerenshaw2003}, which required not only \nuc{13}{C} and \nuc{15}{N} labelling but also selective replacement of one of the two $\textrm{C}_\alpha$ protons by deuterium.

\begin{figure}
\begin{center}
\includegraphics{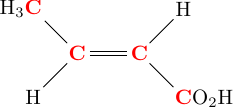}
\end{center}
\caption{Four qubit homonuclear systems are dominated by \nuc{13}{C} labelled crotonic acid.} \label{fig:4qubitcrot}
\end{figure}
Four qubit experiments have, however, become dominated by two systems. The first is an extension of the FFF system iodotrifluoroethylene to make a four spin system by using a \nuc{13}{C} spin \cite{Pan2014, Wang2014b, Peng2014, Zheng2015, Zhou2016, Li2017e, Li2017f, Yu2017, Yao2017, Chang2018, Li2019b, Zhou2019, Luo2019, Wang2020, Zhou2020, Nie2020, Wen2021, Zhao2021a, Chen2021a, Xin2021, Zhang2022, Nie2022}, apparently at natural abundance. The second is the CCCC system provided by fully \nuc{13}{C} labelled crotonic acid (\textit{trans}-but-2-enoic acid) \cite{Knill2000} with \nuc{1}{H} decoupling \cite{Bhole2020, Boulant2002, Xin2017, Wei2014a, Boulant2003, Ryan2005, Passante2009, Ju2010, Wu2011, Lu2016a, Li2017a, Xin2018, Xin2018a, Kong2018, Kong2019, Xin2019, Xin2019a, Li2019, Luo2019a, Wen2019, Wen2019a, Wen2020, Xin2020, Xin2020a, Violaris2021, Li2021}. This system has also been used to implement three qubit experiments by simply choosing only three of the spins \cite{Katiyar2017}, or to implement five to seven qubits by including the \nuc{1}{H} nuclei, as discussed below.

\subsection{Larger spin systems}
Experiments involving more than four spins are much rarer than those involving the small spin systems described above, but a range of larger spin systems has been investigated. An early example was a system of five \nuc{19}{F} nuclei and two \nuc{13}{C} nuclei in a partly \nuc{13}{C} labelled perfluorobutadienyl iron complex \cite{Vandersypen2001} which was used to implement Shor's algorithm to factor 15. More modern experiments however have largely concentrated on crotonic acid, by extending consideration to the \nuc{1}{H} nuclei. These can be divided into three groups: the hydroxyl group, which undergoes rapid exchange and so can be ignored; the two hydrogens attached to C$_2$ and C$_3$ either side of the double bond, which are well suited to use as qubits; and the three hydrogens in the methyl group, which are complicated. These three spins are magnetically equivalent \cite{MHLbook}, and so must be considered together as a group \cite{Corio1967,LyndenBell1969}. The three identical spin-\half\ nuclei can most conveniently be treated as an uncoupled combination of a spin-{\ensuremath{\textstyle\frac{3}{2}}} component and a spin-\half\ component, and the spin-\half\ component can be considered as forming a qubit \cite{Knill2000,Knill1999}. The presence of the spin-{\ensuremath{\textstyle\frac{3}{2}}} component means that this equivalence is not perfect, but it is good enough for some purposes. This allows crotonic acid to be used as a seven qubit system \cite{Knill2000, Knill1999, Zhang2012a, Long2003, Souza2011, Long2022}, although the nature of the methyl hydrogens is sometimes considered to reduce this to a ``six and a half'' qubit system. The same molecule has also been used to implement five qubit experiments by using just the methyl hydrogens and the \nuc{13}{C} nuclei by selecting the \ket{00} component of the other two \nuc{1}{H} nuclei \cite{Knill2001, Ryan2005a}, as discussed in Section~\ref{sec:passive}.

Beyond these heteronuclear systems, homonuclear systems have also been explored. A five qubit system can be implemented using five of the six \nuc{13}{C} nuclei in fully labelled arginine, which form a linear chain that is not significantly coupled to the final carbon in the guanidino group \cite{Wu2011}. A seven qubit system has been demonstrated using all seven \nuc{13}{C} nuclei in a fully labelled cyclobutanone derivative, specifically a racemic mixture of (1S,4S,5S)-7,7-dichloro-6-oxo-2-thiabicyclo[3.2.0]heptane-4-carboxylic acid and its enantiomer \cite{Li2016a, Lu2015, Park2016a}, which has also been used as a six qubit system by ignoring one of the \nuc{13}{C} nuclei \cite{Li2019c}.

This molecule also contains five \nuc{1}{H} nuclei, which are all inequivalent, and so can be used as a twelve qubit heteronuclear system \cite{Lu2017a, Li2019a, Peterson2020}. Twelve qubit experiments have also been demonstrated using \nuc{1}{H}, \nuc{13}{C} and \nuc{15}{N} nuclei in double labelled histidine \cite{Negrevergne2006}. Even larger systems have been studied \cite{Simmons2010, Pande2017, Pal2018} by exploiting star topology molecules \cite{Jones2009, Peng2015, Mahesh2021}, but as these systems do not permit full independent control of the qubits I do not consider them here.

\section{Quantum control} \label{sec:control}
\subsection{Unitary and non-unitary evolution}
The evolution of any purely quantum system under a Hamiltonian is described by the time-dependent Schr\"odinger equation
\begin{equation}
\rmi\,\frac{\partial\ket{\psi}}{\partial t}={\mathcal H}\ket\psi \label{eq:TDSE}
\end{equation}
where natural units have been chosen so that $\hbar=1$, and the Hamiltonian need not be fixed but can vary with time. This has the formal solution
\begin{equation}
\ket{\psi(t)}=U(t)\ket{\psi(0)}
\end{equation}
depending on the propagator
\begin{equation}
U(t)={\mathcal T}\exp\left\{-\rmi\int_0^t {\mathcal H}(t')\,\rmd t'\right\}
\end{equation}
where the Dyson time-ordering operator, $\mathcal T$, defines a procedure for correctly evaluating the operator exponential, as the Hamiltonian at any particular time need not commute with Hamiltonians at other times \cite{EBWbook}. As the Hamiltonian is Hermitian the propagator must be unitary. This means that pure states remain pure, or equivalently that properly normalised ket vectors evolve to other properly normalised kets, and that the inner product between different kets is preserved by the evolution,
\begin{equation}
\inner{\phi(t)}{\psi(t)}=\bra{\phi(0)}U^\dag U\ket{\psi(0)}=\inner{\phi(0)}{\psi(0)},
\end{equation}
since $U^\dag U$ is equal to the identity for any unitary operator. The evolution of a mixed state $\rho$ is given by
\begin{equation}
\rho(t)=U\rho(0)U^\dag,
\end{equation}
and the equivalent result is that unitary evolution does not change the eigenvalues of the density matrix. 

Actually evaluating the propagator for a general time-varying Hamiltonian is only possible in very special cases, but is straightforward when the Hamiltonian is piecewise constant, taking some fixed value ${\mathcal H}_j$ for some time $t_j$. In this case the sub-propagator for any individual time period is
\begin{equation}
V_j=\exp(-\rmi{\mathcal H}_jt_j),\label{eq:Vj}
\end{equation}
where the matrix exponential can be calculated in many different ways \cite{Moler2003}. The combined propagator is given by the time ordered product
\begin{equation}
V=V_n\dots V_2V_1,\label{eq:Vprod}
\end{equation}
with time running from right to left. This structure will be key throughout the following sections.

It might appear from the above that the evolution of a quantum system is always unitary, and this is true if the system is isolated. In reality, however, quantum systems are always coupled to some sort of surrounding environment, and this can lead to effective non-unitary evolution. The evolution of the combination of the system and its surroundings remains unitary, but the evolution of the system alone need not. Formally this occurs because couplings cause the state of the system to become entangled with the state of the surroundings, and performing a partial trace over the surroundings will affect the reduced density matrix describing the state of the system alone \cite{NCbook}.

The most obvious type of non-unitary evolution is relaxation, which arises from uncontrolled couplings to the environment. Relaxation can be broadly divided into decoherence, or dephasing (transverse relaxation), which acts to remove off-diagonal elements from the density matrix, and longitudinal relaxation, which changes the diagonal elements, driving them towards the thermal equilibrium state. In conventional NMR decoherence, which occurs with a time constant $T_2$, is a bad thing in that it limits resolution and sensitivity, although measurements of decoherence rates can be used to extract information on molecular motion \cite{Palmer2004}, but longitudinal relaxation, which occurs with a time constant $T_1$, is essential to produce the initial population differences that lead to detectable signals. Within QIP, however, all forms of relaxation are unambiguously a bad thing, as they introduce errors into the quantum state, which must either be resisted (using decoherence free subspaces \cite{Zanardi1997, Duan1997, Lidar1998, Lidar2001, Lidar2001a}) or detected and correction (quantum error correction \cite{Shor1995, Steane1996, Calderbank1996, Laflamme1996}). State preparation in technologies other than NMR is usually performed using some explicit reset mechanism, such as optical pumping, rather than relying on natural relaxation to a thermal state.

While uncontrolled evolution is a bad thing, controlled non-unitary evolution does have uses in QIP. The most important example is projective quantum measurement, which in effect causes a superposition to collapse into an eigenstate. As well as being needed to extract a definite result from an algorithm which ends in a superposition state this provides a simple route to reset qubits, such as ancilla qubits used in quantum error correction, permitting them to be reused. Unfortunately, projective measurements are not available in conventional NMR. Instead, the most important non-unitary operations available are magnetic field gradients and phase cycling.

Field gradients \cite{Barker1985} cause the Larmor frequency, and thus the evolution, to vary over the macroscopic sample. As the detection process combines signals from all over the sample the effect is to observe an average density matrix. For this reason the process is normally referred to within QIP as \textit{spatial averaging}. In effect the evolution of a particular molecule becomes entangled with its position, and the position is then ``traced out'' by simultaneous detection of the whole sample \cite{Pravia1999}, which is equivalent to performing a partial trace over the position label. The result is similar to imposing a decoherence process on the system, but with two significant differences. Firstly, zero-quantum coherences are invulnerable to gradients in homonuclear systems: this natural example of a decoherence free subspace can sometimes be useful \cite{Hall1986}, but is more frequently a problem \cite{Macura1981}. Secondly the dephasing can be reversed in spin echoes, allowing the dephasing to be applied selectively to some spins and not others. The effectiveness of spin echoes is reduced by diffusion \cite{Stejskal1965}, and this provides a convenient route to controllable decoherence \cite{Cory1998, Fitzsimons2015}.

Phase cycling is a major topic in conventional NMR, but in principle it simply refers to performing an experiment several times with different phases for some pulses, and then combining the results together, with the intention of retaining some desired signals while cancelling others \cite{Bodenhausen1977, Bain1984, Bodenhausen1984}. Within quantum information processing this is normally called \textit{temporal averaging}, as the evolution is averaged over experiments performed at different points in time. Temporal averaging can be generalised to include experiments that differ in other ways \cite{Knill1998}, but as in conventional NMR the cleanest results are obtained when the experiments are most similar to one another, and phase cycling remains a common approach. Unlike the use of field gradients phase cycling can discriminate between nuclear species, and can be applied to individual spins by using selective pulses, thus permitting the suppression of zero-quantum terms. The simplest approach, exhaustive temporal averaging, can become extremely long, but it may be sufficient just to select a subset of experiments \cite{Bhole2020, Knill1998}.

I will consider non-unitary processes again in Section~\ref{sec:PPS}, which discusses pseudo-pure states, but until then will concentrate on unitary transformations.

\subsection{Quantum logic gates}
One central task in implementing QIP is to implement quantum logic gates. Fundamentally these are just unitary transformations whose action on quantum bits has a simple interpretation in terms of information processing. A wide range of notations are used, but they all represent the same basic operations. As these operations are unitary it suffices to write down a unitary matrix which has the desired effect. The simplest example is the X gate, which converts the basis state \ket{0} to \ket{1}; the reason for referring to this operation as X will soon become clear. This gate is described by the unitary propagator
\begin{equation}
\mathrm{X}=\begin{pmatrix}0&1\\1&0\end{pmatrix},\label{eq:Xgate}
\end{equation}
which is easily shown to have the desired effect, as
\begin{equation}
\mathrm{X}\ket{0}=\begin{pmatrix}0&1\\1&0\end{pmatrix}\begin{pmatrix}1\\0\end{pmatrix}=\begin{pmatrix}0\\1\end{pmatrix}=\ket{1},
\end{equation}
and so on.

The X gate has a simple action on the basis states and so a simple interpretation in terms of classical information processing, implementing the \NOT\ operation. However, as X is a unitary propagator it can also be applied to superposition states, since
\begin{equation}
\begin{split}
\mathrm{X}(c_0\ket0+c_1\ket1)
&=c_0\mathrm{X}\ket0+c_1\mathrm{X}\ket1\\
&=c_0\ket1+c_1\ket0
\end{split}
\end{equation}
by linearity. If the initial and final states are viewed on the Bloch sphere, as described in Section~\ref{sec:states}, then the action of X is to rotate the state around the $x$-axis by $180^\circ$, explaining the name. In the same way the Z gate,
\begin{equation}
\mathrm{Z}=\begin{pmatrix}1&0\\0&-1\end{pmatrix},
\end{equation}
acts to rotate the state around the $z$-axis by $180^\circ$. Unlike X this gate has no classical interpretation, but is a purely quantum logic gate. Another purely quantum gate is the Hadamard gate,
\begin{equation}
\mathrm{H}=\frac{1}{\sqrt2}\begin{pmatrix}1&1\\1&-1\end{pmatrix},
\end{equation}
which interconverts basis states and superpositions.

The matrices describing all these gates are unitary, which is easily shown by direct calculation, and so these gates correspond to possible unitary propagators, and can in principle be implemented by evolution under some Hermitian Hamiltonian. In most cases the required Hamiltonian will not be immediately available, and so it will be necessary to achieve the desired unitary evolution by combining a number of steps. In the language of NMR it is possible to construct an \textit{average Hamiltonian} corresponding to the desired evolution, although his language is rarely used within QIP, where it is more normal to think about the propagators rather than the Hamiltonian. One exception to this general rule is the use of refocusing sequences, explored in Sections \ref{sec:WHSE} and \ref{sec:dd}.

\subsection{The control problem}
Although a wide range of different approaches have been explored for controlling NMR implementations of QIP, at heart they all have the same structure \cite{Arenz2017}. The system has a background Hamiltonian, $\mathcal H_0$, sometimes called the \textit{drift} Hamiltonian, which describes the free evolution of the system and contains Zeeman and spin--spin coupling terms. The NMR spectrometer can then be used to apply additional \textit{control} Hamiltonians, which are RF fields, usually at single frequencies near resonance with one or more spins. The overall evolution of the quantum system is controlled by varying the control Hamiltonians, by changing the RF amplitude, phase, and in some cases frequency.

Unitary control is relatively straightforward in a fully heteronuclear system. Each spin can be viewed on resonance in its own rotating frame, so that the free evolution only involves the couplings, which are usually quite small in comparison with easily achievable RF nutation rates. In this case it is a reasonable approximation to simply ignore the drift Hamiltonian during briefly applied pulses of the control Hamiltonians. With separate control of amplitude and phase at the resonance frequency for each spin it is easy to apply any desired single-qubit gate, while two-qubit gates can be implemented using free evolution under the couplings, most simply by using spin echoes to construct controlled-phase gates \cite{Jones1998b}. This provides a universal set of quantum logic gates \cite{Barenco1995} and so any desired evolution can be approximated to arbitrary accuracy, and by the Solovay--Kitaev theorem this can be done efficiently \cite{Dawson2006}. Practical methods for the design of efficient refocusing networks will be discussed in Section~\ref{sec:WHSE}.

The situation is more complex with homonuclear spin systems. Fundamentally this is because the spin-selective shaped pulses \cite{Freeman1998} necessary to perform qubit-selective gates have a minimum length, set by the smallest frequency gap between the resonance frequencies of different spins, and so it is necessary to consider evolution under the full Hamiltonian, combining drift and control terms. The first homonuclear implementation of a quantum algorithm \cite{Jones1998c} involved two \nuc{1}{H} spins, with frequency selection achieved using Gaussian shaped pulses \cite{Bauer1984}, incorporating a phase ramp to move the resonance frequency between the two spins \cite{Hedges1988, Boyd1989}. Choosing the pulse length to be stroboscopic with the frequency difference between the two spins means that the total evolution experienced by the other spin under its Zeeman Hamiltonian corresponds to an integer number of rotations and can be ignored \cite{Jones1998c}. An alternative approach is to use jump and return sequences \cite{Anwar2004, Jones1999a, Plateau1982}, which achieve selective excitation in the shortest possible time \cite{Bowdrey2006}.

This stroboscopic approach only works for two spins, however, and beyond this it is becomes challenging to use conventional shaped pulses as it becomes necessary to worry about the phase of every spin. The most direct approach, sometimes called \textit{abstract reference frames} \cite{Knill2000}, simply creates a virtual transmitter for each spin, using conventional phase ramped pulses, which are kept phase coherent with the resonant spin. Unlike in heteronuclear systems, these pulses will weakly affect off-resonant spins through transient Bloch--Siegert shifts, but it is possible to calculate the sizes of these shifts and offset the abstract reference frames appropriately. These calculations are conveniently combined with a pulse sequence compiler \cite{Ryan2008} which keeps track of phases. A similar approach can be used to track extraneous spin--spin couplings, to avoid unnecessary refocusing operations \cite{Bowdrey2005}.

A more direct approach is to replace conventional selective pulses, which avoid exciting unselected spins but do not leave them truly unchanged, with more sophisticated pulses which perform an identity operation on the unselected spins. In this case there are no phase errors to keep track of, but it is no longer possible to design pulses using simple intuitive methods. Instead it is necessary to use methods such as \textit{optimal control theory} to find pulses with the correct behaviour \cite{Suter2008, Mahesh2022, Kuprov2023}. While such methods are intrinsically far more complex than conventional pulse designs, the fact that it is only necessary to obtain the desired behaviour at a small number of distinct frequencies, which are known at the start of the process, provides a useful simplification.

\subsection{Global phases}
Global phases arise in quantum mechanics because the conventional description of a quantum state in terms of a ket contains more information than the state itself does. They are rarely a concern in conventional NMR because the use of notations based on density matrices causes them to disappear. This is obvious for a pure state density matrix, since
\begin{equation}
\rho'=\ket{\psi'}\bra{\psi'}=\rme^{\rmi\gamma}\ket\psi\bra\psi\rme^{-\rmi\gamma}=\ket\psi\bra\psi=\rho,
\end{equation}
where the two global phases are just scalars, and so can be moved to the same side, where they obviously cancel. Mixed states are simply averages over pure states, and so the same argument applies, and this generalises to NMR operators such as $I_z$. Such operators are not really density matrices (in particular they have trace equal to zero, while all properly normalised density matrices have trace equal to one) and within NMR QIP they are usually called \textit{deviation density matrices} \cite{Gershenfeld1997}, but their behaviour towards global phases is identical to that of true density matrices.

Global phases are also an issue when considering propagators, and here can cause more serious concerns. For any propagator $U$ there is an infinite family of equivalent propagators,
\begin{equation}
U'=\rme^{\rmi\gamma}U,
\end{equation}
whose action on a ket differs only by a physically irrelevant global phase, which cancels out for density matrices as usual. Thus $U$ and $U'$ are entirely equivalent, but they are not actually identical.

Different but equivalent propagators must correspond to different but equivalent Hamiltonians, and global phases arise from elements in the Hamiltonian which are proportional to the identity operator. Equivalently, different global phases correspond to different positions for the zero point of the energy scale, which have no physical significance as only energy differences are physically meaningful. Such terms do not arise in conventional NMR treatments, where all Hamiltonians are combinations of traceless operators, but they are regularly seen in other physical systems, where the energy zero is frequently placed at the energetic ground state, rather than at the zero-field spin energy as done in NMR.

Problems with global phases will not normally arise if a consistent notation is used throughout, but problems can arise when combining, for example, NMR notation with theoretical QIP notation. Most of the fundamental logic gates used in QIP do not correspond to traceless Hamiltonians, and within NMR QIP can only be implemented with a global phase shift. For example the \NOT\ gate is implemented as a $180^\circ_x$ rotation, but this has the propagator
\begin{equation}
\exp(-\rmi\pi I_x)=\begin{pmatrix}0&-\rmi\\-\rmi&0\end{pmatrix}
\end{equation}
which differs from the desired X gate (equation~\ref{eq:Xgate}) by a global phase of $-\rmi$. When seeking to implement a \NOT\ gate in NMR it is essential either to use a fidelity measure that ignores global phase differences, or to ensure that the target has the appropriate global phase.

This second approach can largely be achieved by specifying targets in NMR notation, but even then a subtlety can arise: spin-\half\ particles exhibit spinor behaviour \cite{Pravia1999}, and thus pick up a global phase of $-1$ on being rotated through a full circle. Thus the operators for a $180^\circ_x$ and a $540^\circ_x$ rotation differ by a sign, even though they have identical physical effects, and the same is true for $180^\circ_x$ and $180^\circ_{-x}$ rotations. This phenomenon is important in, for example, the design of composite pulses, where the global phase in the target unitary may have to be allowed for \cite{Jones2013}.

\section{Optimal control}\label{sec:OC}
The basic idea of optimal control \cite{Warren1993, Brif2010} is to use numerical searches to locate a set of time-varying controls which optimally implements some desired unitary transformation $U$ in the presence of a fixed drift Hamiltonian. The overall Hamiltonian
\begin{equation}
{\mathcal H}(t)={\mathcal H}_0+{\mathcal H}_1(t)
\end{equation}
is best considered in some suitable rotating frame where it can be taken as piecewise continuous, permitting the corresponding unitary transformation $V$ to be calculated using equations \ref{eq:Vj} and \ref{eq:Vprod}. From this a transformation fidelity can be calculated as
\begin{equation}
{\mathcal F}=\left|\frac{{\rm tr}(U^\dag V)}{{\rm tr}(U^\dag U)}\right|^2.\label{eq:Ufid}
\end{equation}
Note that if $V=U$ then $U^\dag V$ is equal to the identity, and so the trace is maximised; taking the square of the absolute value removes any global phase differences, while the denominator acts to normalise the result into the range $0\le{\mathcal F}\le1$, as desired for a fidelity measure. The task is then to locate a parameterised set of values of ${\mathcal H}_1(t)$ which maximises ${\mathcal F}$. The parameterisation can be as simple as the strengths of the control Hamiltonians at each point in the piecewise continuous form, or can be more complex and indirect. To better reflect experimental limitations it may prove necessary to restrict the strengths of control fields, or at least to penalise solutions which require unrealistically strong fields, and it can also be useful to seek solutions whose fidelities are robust with respect to minor errors in the drift and control Hamiltonians.

Within this general class of problems many different approaches have been explored. These vary principally in the choice of optimization algorithm, the choice of fidelity measure, and any restrictions that are placed on the form of ${\mathcal H}_1(t)$, as briefly outlined below.

\subsection{Optimization algorithms}\label{sec:OA}
Since the quality of a chosen set of controls is summarised by a real number, the fidelity, optimisation can be performed using any general-purpose algorithm to maximise the fidelity, or equivalently to minimize the \textit{infidelity}, defined by
\begin{equation}
{\mathcal I}=1-{\mathcal F}.\label{eq:infid}
\end{equation}
A wide range of minimization algorithms are available, but these can be divided into broad categories according to the use that the algorithm makes of gradients, and any measures that the algorithm takes to guard against becoming trapped in local minima.

Perhaps the simplest approach is the simplex algorithm \cite{Nelder1965}, which seeks a local minimum in an $n$-dimensional search space by exploring $n+1$ distinct points. The function is evaluated at each of the points forming the vertices of this \textit{simplex}, and an attempt is made to improve the current worst point by a series of operations which move it in the general direction of the better points. Eventually the simplex will surround a local minimum, and will then contract so that all the vertices approximately coincide at the minimum. This approach requires only that the function can be evaluated at any point, and in particular the function does not need to be differentiable. It is also relatively robust to situations where the function cannot in fact be precisely evaluated, but only estimated to within some uncertainty. This can be relevant in the case of \textit{closed-loop control}, discussed in Section~\ref{sec:CLC}, where the fidelity is determined experimentally rather than evaluated computationally, and the uncertainty is governed by noise. A simple example familiar from conventional NMR is provided by computer adjustment of shim coil currents to maximise the size of a deuterium lock signal \cite{Ernst1968}.

More rapid convergence can normally be achieved if the algorithm has access to gradients of the function with respect to the control parameters. (The use of $n+1$ distinct points means that the simplex algorithm has implicit access to gradient information through finite differences \cite{Scheinberg2022}, but the gradients are not explicitly calculated or used.) The most obvious approach, \textit{steepest descent}, simply moves in the direction of the gradient until the value of the function stops decreasing. This method is ancient \cite{Cauchy1847, Lemarechal2012} but converges less rapidly than a naive consideration might suggest. To obtain more rapid convergence it is necessary to use a method such as \textit{conjugate gradients} \cite{Mao1986, Shewchuk1994} which avoids the zig-zag paths imposed by steepest descent. Even better convergence is obtained by using the Hessian, that is the matrix of second derivatives of the function, but finding this may be rather tedious if a large number of control variables are involved. An excellent compromise is provided by the second order quasi-Newton Broyden--Fletcher--Goldfarb--Shanno (BFGS) algorithm \cite{Broyden1970, Fletcher1970, Goldfarb1970, Shanno1970}, which approximates the Hessian using values of the gradients from successive steps, and so gives rapid convergence near the minimum without excessive overhead in the earlier stages \cite{DeFouquieres2011}. The BFGS algorithm is available in many standard mathematical packages: for example the Matlab minimisation function \texttt{fminunc} has options to use both conventional BFGS and the limited memory L-BFGS variant \cite{Liu1989, Byrd1994} when gradients are provided.

All of these algorithms will converge on some minimum, but a function may possess multiple minima, and the aim is to find the lowest (or equal lowest) of these, which is a global minimum. In the most general case it is very hard to be sure that this has been achieved, but there are several approaches for tackling the problem. Most simply, if the infidelity function is confined to lie between 0 and 1 then any minimum with an infidelity equal to 0 must be a global minimum, and pragmatically any point with a sufficient small infidelity is good enough. If the search algorithm converges to a point which is not good enough, then the search can be restarted from a different initial position, with the hope of converging on a better local minimum.

A more sophisticated alternative is provided by \textit{simulated annealing} \cite{Kirkpatrick1983}, which builds on the earlier Metropolis algorithm \cite{Metropolis1953}. While conventional minimization algorithms only ever move downhill a simulated annealing algorithm may also move uphill, just as thermal excitations can allow a physical system to cross an energy barrier to reach a lower energy state. As the algorithm progresses the equivalent temperature of the process, which determines the probability of accepting an uphill move, is gradually reduced, so that the algorithm turns smoothly into a conventional minimization process. The method is particularly effective at locating a deep global minimum surrounded by shallow local minima, but provides no protection against the presence of multiple deep but suboptimal minima, as the algorithm is likely to become trapped by the first deep minimum that it finds. A closely related algorithm, threshold acceptance, can perform the same search more rapidly \cite{Franz2001}, but does not overcome the fundamental problem of deep but false minima.

Simulated annealing was swiftly applied to the problem of NMR pulse design \cite{Morris1989, Geen1989a}, most famously in the development of the BURP (Band-selective, Uniform Response, Pure-phase) family of pulses \cite{Geen1991}. The threshold acceptance algorithm has been used in combination with more conventional minimization to design control pulses for NMR QIP \cite{Ram2022}. Note that simulated annealing should not be confused with quantum annealing \cite{Finnila1994, Kadowaki1998}, which seeks to minimise a function using explicitly quantum hardware \cite{Johnson2011}, and has been demonstrated using an NMR implementation \cite{Chen2011a}.

A quite different approach is provided by \textit{genetic algorithms} \cite{Whitley1994}, also known as evolutionary algorithms. A set of controls can be considered as a genotype, with the corresponding unitary transformation being the phenotype, and the fidelity providing a fitness function, which should be maximized. The algorithm begins with a random selection of genotypes, from which the fittest members are selected. New members are then generated by a combination of mutation and crossing existing members, and the process is repeated. The process of selection means that the highest fidelity sequences will be retained, while the mutation and crossing processes allow the control space to be explored.

Although the genetic approach appears promising, in practice it is only useful when the parametrisation of the problem means that new population members retain some common features with their antecedents. In other cases mutation and crossing effectively produce entirely unrelated trial solutions, and the genetic algorithm becomes simply a complicated way of optimising a function by sampling values at random.  After an early application in designing shaped pulses \cite{Wu1989}, interest within conventional NMR largely moved to its use in automated analysis \cite{Cho2008, Schleif2010}, but more recently the technique has been applied to solid state NMR \cite{Manu2015, Manu2016}, to \textit{in vivo} NMR \cite{Somai2022}, and to NMR QIP \cite{Devra2018, Bhole2016, Manu2012}.

\subsection{Fidelity measures}
The fidelity measure introduced above, equation~\ref{eq:Ufid}, is not the only possible choice. One apparently obvious alternative is to take the square root of this definition, using the absolute value of the trace rather than its square. This is a simple monotonic transformation, and so the choice may seem arbitrary, but taking the square has practical advantages. In particular, evaluating the differential of an absolute value is messy, while its square is much better behaved, since $|y|^2=y^*y$ leads to
\begin{equation}
\frac{\rmd |y|^2}{\rmd x}
=y^*\frac{\rmd y}{\rmd x}+y\frac{\rmd y^*}{\rmd x}
=2\,{\mathrm{Re}}\left(y^*\frac{\rmd y}{\rmd x}\right),\label{eq:dy2}
\end{equation}
where Re indicates taking the real part. This form also shows clearly that when the fidelity is close to unity then taking the square root halves the calculated infidelity.

One important exception to this occurs when the global phase difference between $U$ and $V$ is known beforehand, a situation which can occur in the design of simple composite pulses \cite{Jones2013}. In this case the global phase can be corrected before calculating the fidelity, and there is no need to take an absolute value. This leads to the simplest possible fidelity and gradient functions, which is particularly useful when analytic methods are used. For numerical optimisation, however, the robustness of taking the square modulus makes it the simplest and most straightforward approach.

A far more significant change is to replace this propagator fidelity with a state fidelity, such as
\begin{equation}
{\mathcal F}_\psi=|\bra\psi U^\dag V\ket\psi|^2=\bra\psi U^\dag V\ket\psi\bra\psi V^\dag U\ket\psi\label{eq:psifid}
\end{equation}
which measures how accurately $V$ changes $\ket\psi$ into the desired state $U\ket\psi$. Such state-to-state fidelities (also called point-to-point fidelities) are frequently used in conventional NMR, but are only rarely used in QIP, as the initial state before applying a logic gate is not normally known. The two forms can be related by averaging the state-to-state fidelity over a sufficiently wide range of input states, and this approach is particularly useful for single-qubit gates \cite{Bowdrey2002}, where it suffices to average over three states corresponding to the cardinal axes of the Bloch sphere.

An apparent computational advantage of this approach is that the state $V\ket\psi$ can be obtained by numerical integration of equation~\ref{eq:TDSE} without the need to explicitly determine the propagator $V$ \cite{Kuprov2023}, which is a key feature of approaches such as Spinach \cite{Hogben2011, Allami2019}. However, finding sequences that perform the correct unitary transformation requires averaging over a large number of input states, and the expense of doing so wipes this gain out. For conventional NMR there is much to be said for following the straight and narrow path: ``do not open krons, do not diagonalise, use cheap norm estimators, and do not exponentiate matrices'' \cite{Kuprov2019}, but for QIP it is vital to remember the caveat ``unless you absolutely have to'' \cite{Kuprov2019}.

A second advantage of state-to-state fidelities is that they can be generalised to non-unitary evolution. Equation~\ref{eq:psifid} can be rewritten as
\begin{equation}
{\mathcal F}_\rho=\bra\psi U^\dag \rho\, U\ket\psi\label{eq:rhofid}
\end{equation}
where
\begin{equation}
\rho=V\ket\psi\bra\psi V^\dag
\end{equation}
is the density matrix corresponding to the pure state $V\ket\psi$, but the same fidelity equation can be used when $\rho$ is a mixed state density matrix, arising from $\ket\psi$ by some more general process. This form enables the design of optimal state transfers \cite{Khaneja2005} and quantum gates \cite{SchulteHerbrueggen2011} in the presence of significant relaxation processes.

It is tempting to generalise this formula even further, and to define a fidelity between two density matrices $\rho$ and $\sigma$ as something like ${\textrm{tr}}(\rho\sigma)$, but as discussed in Section~\ref{sec:states} this form is only correct if at least one of the density matrices corresponds to a pure state. Instead it is necessary to use the Uhlmann--Jozsa fidelity, as discussed in Section~\ref{sec:fidelities}. In addition to being somewhat complex to calculate, even using the efficient form \cite{Baldwin2023}, the interpretation of any mixed state fidelity can be unintuitive. For now I will simply ignore the question, although I will return to it in Section~\ref{sec:PPS}.

Despite the warning above, the naive fidelity expression ${\textrm{tr}}(\rho\sigma)$ is very frequently used in conventional NMR, and is commonly extended to calculations involving deviation density matrices such as product operators. Such expressions are not normally genuine fidelities, and in particular are not normally restricted to values between 0 and 1, but they can provide a useful and easily calculated function to maximise or minimise. Fortunately, these naive expressions can be used when comparing one density matrix with different unitary transformations of another density matrix \cite{Glaser1998}, and this is frequently sufficient. It is necessary to be careful when extending this definition to operators describing coherence orders rather than magnetizations \cite{MHLbook}, as these are not Hermitian and so not equal to their adjoints, and it is necessary to distinguish carefully between ${\textrm{tr}}(\rho\sigma)$ and ${\textrm{tr}}(\rho^\dag\sigma)$.

The optimization function can also be used to design pulses subject to specific constraints by adding a penalty function which discourages, for example, large control amplitudes or rapid changes in amplitude \cite{Goodwin2018}. Such mixed optimization functions are not strictly speaking fidelities, but their behaviour can be very similar, particularly if penalties are only applied above some threshold. However it is generally better to avoid the use of penalty functions if their aim can be achieved in some other way \cite{Rowland2012}, such as the restricted forms discussed below. When such mixed optimization functions are used it is important to be aware that the ``fidelity'' might not be confined to the conventional range of 0 to 1, and so the infidelity calculation in equation~\ref{eq:infid} is not always appropriate.

\subsection{Robustness to errors}
Until now I have assumed that the control Hamiltonian experienced by a particular spin system is equal to the control Hamiltonian that was nominally applied, but in practice this will not be the case \cite{Boulant2003, Pravia2003}. Although the strengths of control fields can be calibrated by simple measurements, the assumption that a control field has a fixed strength is incorrect. The NMR sample is macroscopic and the applied RF field will vary significantly over the sample. The exact pattern of $B_1$ inhomogeneity will depend on the sample and the RF coil, but in a typical NMR system the main distribution is approximately Gaussian, with a width of around $\pm5$\%, and a significant tail at much lower values \cite{Husain2013}. This can be reduced by using a small sample \cite{Skinner2004}, or by using NMR methods to select regions of high homogeneity \cite{Knill2000, Souza2011, Cory1993}, but cannot be entirely eliminated, and is a particularly serious problem with early designs of cryogenic probes \cite{Kobzar2004}. Errors can also arise if the $B_1$ field strength is miscalibrated, or if it changes after calibration, for example due to temperature changes in the RF amplifier.

Tackling $B_1$ strength errors is a major topic in conventional NMR, notably through the use of composite pulses \cite{Levitt1979, Levitt1986}, and is also an important topic in NMR QIP. Such systematic errors can be addressed because they are reproducible, and so can be arranged to largely cancel out. Fortunately it is easy to build a requirement for robustness into optimal control by simply averaging the fidelity over a range of different control field strengths \cite{Khaneja2005}, although more sophisticated processes have also been considered \cite{Kosut2022, Laforgue2022}. It is not normally necessary to choose this range particularly carefully or to sample the range finely, and choosing field strengths such as 97\%, 100\% and 103\% of the nominal value seems to work well in practice. The variation of fidelity with field strength is usually slow enough that a pulse that performs well at these three values will perform adequately across the whole of the main part of the distribution. Dealing with spins in the tail of the distribution, with very low $B_1$ strengths, is far more challenging, and rarely worth the effort.

With a heteronuclear spin system it is important to remember that the RF field inhomogeneity pattern may be different for different nuclei. A typical NMR probe has two physical coils, an inner coil with a high filling factor \cite{Hoult1976} and a larger outer coil. Each coil may be tuned to multiple resonance frequencies \cite{Mispelter2015}, most commonly placing high frequencies on one coil and low frequencies on the other; ideally the nucleus actually detected should be placed on the more sensitive inner coil, with the outer coil only used to apply control fields, but it is common, if not ideal, for QIP experiments to be performed on systems optimised for other conventional purposes, and so for probes to be used the wrong way round. The RF field inhomogeneity depends strongly on the coil geometry, and only weakly on the RF frequency, and so will be very similar for all nuclei addressed through the same coil. It is therefore sufficient to consider at most two sets of field distributions, and so average over nine combinations of different field strengths.

A further problem can arise when the RF field strength varies \textit{during} a control pulse, for example if the power of an RF amplifier rises or falls after it has been activated \cite{Skinner2004}, or as a consequence of the finite response times of tuned circuits \cite{Rasulov2023}. While some cases can be modelled fairly accurately, the most general errors have to be addressed in another way, such as monitoring the RF amplitude during a pulse using a pickup coil \cite{Ryan2009}, or by using closed-loop control.

In early work it was common to design control sequences to be robust to other types of systematic error, such as variations in the chemical shift. In practice this is usually unnecessary for QIP, and RF inhomogeneity is normally the only important effect to consider. This is very different from the situation in conventional NMR, where the use of optimal control theory to design band-selective pulses is a very important topic, and this is addressed briefly in Section~\ref{sec:SSC}. An interesting modern exception to this general rule is the design of sequences which are robust to the spin states of passive spins, a point explored in more detail in section~\ref{sec:passive}.

\subsection{General and restricted forms}
Optimal control requires finding a set of control fields that achieve a desired aim, and it is important to consider how these control fields are parameterised. I am assuming that the fields will be piecewise continuous, to enable a practical solution of equation~\ref{eq:TDSE}, and the simplest approach is just to digitise the control fields at equally spaced intervals in time, as is normally done when specifying a shaped pulse. For a homonuclear spin system all qubits are affected by the same control field, and so the Hamiltonian is conveniently parameterised as
\begin{equation}
{\mathcal H}_j={\mathcal H}_0+\alpha^x_j F_x +\alpha^y_j F_y\label{eq:Hxy}
\end{equation}
where ${\mathcal H}_0$ includes resonance offset terms and couplings, $\alpha^x_j$ and $\alpha^y_j$ are real amplitudes, and
\begin{equation}
F_x=\sum_k I^k_x,\quad F_y=\sum_k I^k_y
\end{equation}
are the total angular momentum operators across all spins. Alternatively the real amplitudes can be packed together to form a single complex amplitude, $\alpha=\alpha^x+\rmi\alpha^y$, and this can be described using its magnitude and phase rather than its components. In a heteronuclear spin system there are separate control fields, and thus separate amplitudes, for each homonuclear subset of spins.

To access the full flexibility offered by arbitrary control fields it might seem best to sample the control fields as finely in time as possible, but this is not the case. The physical apparatus used to generate the control fields will always have some limiting time resolution, but even above this limit it may prove difficult to actually implement very rapid variations. The analogue parts of any NMR system will always act as low-pass or band-pass filters, smoothing the applied waveform, but more seriously the digital control circuitry can introduce significant switching transients at every change in complex amplitude. This is rarely a major problem with modern spectrometers built around direct digital synthesis \cite{Momo1994, Liang2009}, but imperfections can be very serious for older systems which use switchable attenuators, where much better experimental results are seen with a coarser time spacing \cite{Rowland2012}. Beyond these experimental issues, designing a more finely sampled pulse will clearly require more computer power \cite{Schirmer2011}. A more careful analysis is attempted below, but it is clearly desirable not to sample much more finely than necessary. Fortunately, simple Fourier considerations indicate that a very fine sampling is not normally required.

A common approach is to vary both the $x$ and $y$ components of the control fields, or equivalently to vary both their amplitude and phase, but it can be useful to consider more restricted forms. In particular it can be very convenient to use a fixed amplitude for the control fields and vary only the phase. This avoids any need to impose an amplitude penalty, but also has computational advantages, as discussed in section~\ref{sec:phaseonly} below. A less common approach is to fix the phase and vary only the amplitude, or to use a single control field along $x$, fixing $\alpha^y=0$, which corresponds to restricting the phase to $0$ and $\pi$. This has the disadvantage that any such pulse cannot distinguish between spins at positive and negative values of the same absolute offset frequency.

Several more restrictive approaches have been explored in detail, some of which have counterparts in conventional NMR, and all of which are designed to describe a long shaped pulse with a relatively small number of parameters. One approach is to split a sequence into fixed amplitude pulses and variable length delays, which within QIP is known as \textit{quantum bang--bang} control \cite{Morton2006a}. This approach has been widely explored for dynamical decoupling (Section~\ref{sec:dd}), but also for more general control \cite{Bhole2016, Khurana2017}. At the other extreme some authors have aimed to design smooth pulses by describing the amplitudes in terms of low frequency Fourier components, as seen in conventional NMR in the BURP family of pulses \cite{Geen1991}, and which was more recently applied in NMR QIP \cite{Peterson2020}. In this case the low frequency description is often converted to a high frequency sampled waveform before calculating the evolution, which can cause complications in calculating gradients.

\subsection{Composite pulses}\label{sec:compul}
Another approach of considerable historical importance is \textit{strongly modulating composite pulses} \cite{Fortunato2002, Boulant2003}. Like conventional composite pulses, these construct a shaped pulse from a small number of pulses placed back-to-back, but in addition to the phases the amplitudes, lengths, and offset frequencies are also varied. By using a sequence of frame transformations it is possible to directly calculate the overall evolution in an efficient manner, and for systems with small numbers of qubits excellent single-qubit gates can be designed with ease. The final optimised sequence is then converted to a conventional finely sampled shaped pulse, using phase ramping to implement any frequency shifts \cite{Hedges1988, Boyd1989}.

Strongly modulating pulses have been widely applied in NMR QIP experiments \cite{Fortunato2002, Weinstein2004, Fitzsimons2007, Mitra2008, Du2007, Boulant2003, Negrevergne2006, Pravia2003, Cappellaro2005, Anwar2005, Hodges2007} including solid state \cite{Baugh2005, Baugh2006} and strongly coupled systems \cite{Mahesh2006}, quadrupolar nuclei \cite{Teles2007, Kampermann2005} and ENDOR \cite{Hodges2008}.  For some time it seemed likely that the approach would become the dominant method for designing pulses for NMR QIP, but it has now been effectively superseded by the more general GRAPE technique described in Section~\ref{sec:GRAPE}.

Conventional composite pulses, in which the individual pulses have a fixed common frequency, usually have a fixed common amplitude, and frequently have either a fixed common length or individual fixed lengths which are small multiples of some underlying basic length, are rarely useful for qubit selective addressing. Superficially they appear suitable for use in heteronuclear QIP systems, but even in this case there can be issues arising from evolution under spin--spin couplings when pulses are applied simultaneously to two or more spins, and it may be better to use simple pulses \cite{Xiao2005}. They have, however, found wide application in dynamical decoupling, as described in Section~\ref{sec:dd}, and have also been used in two-qubit homonuclear spin systems, where their tolerance of off-resonance errors permits uniform excitation of both spins \cite{Cummins2001}. This uniform excitation can be combined with jump and return sequences to provide frequency selection \cite{Bowdrey2006}. These applications have led to considerable interest in designing composite pulses for NMR QIP, some of which may have wider applications in conventional NMR. These novel pulses are all universal rotors, which perform well for any initial state, sometimes called Class A composite pulses \cite{Levitt1986}.

The design of robust \NOT\ gates turns out to be much simpler than the more general case, particularly when these gates are made from sequences of $180^\circ$ pulses \cite{Jones2013}. Early results from conventional NMR include the three-pulse sequence $180_{120}\,180_{240}\,180_{120}$, which corrects $B_1$ strength errors, and the related sequence $180_{60}\,180_{120}\,180_{60}$, which tackles off-resonance errors \cite{Tycko1985, Odedra2012b}. (When designed for use in conventional NMR it is common not to try to design a \NOT\ gate but simply to implement a $180^\circ$ rotation around some axis in the $xy$-plane, but this can be easily fixed by offsetting all the phases, and sequences listed here correspond to the desired $180_x$ rotations, up to a global phase of $\pm1$.) A key result is a simple five-pulse sequence
\begin{equation}
180_{240}\,180_{210}\,180_{300}\,180_{210}\,180_{240}
\end{equation}
which tackles both $B_1$ strength and off-resonance errors. Within NMR QIP this is generally called the Knill pulse and is widely used in dynamical decoupling \cite{Souza2011b, Ryan2010}, as discussed in Section~\ref{sec:RDD}. The performance can be further improved with sequences of seven or nine pulses \cite{Jones2013}.

For dealing with $B_1$ strength errors in quantum gates corresponding to other rotation angles, the BB1 sequences designed by Wimperis \cite{Wimperis1994} have proved particularly useful. These provide good suppression of $B_1$ strength errors at no cost to the sensitivity to off-resonance effects, and are available for all pulse flip angles. One minor change when applying them to NMR QIP is that the correction sequence, comprising four $180^\circ$ pulses, is usually placed in the middle of the main error-prone pulse, rather than before it as in Wimperis's original design. For the design of \NOT\ gates ($180^\circ_x$ pulses), the Wimperis sequence can be applied iteratively \cite{Jones2013a}, permitting sequences with arbitrary suppression of $B_1$ strength errors to be designed with relative ease. For other rotation angles this iterative approach is not successful, but a mixture of analytic and numerical searches have found some BB1 style sequences which outperform the classic design \cite{Husain2013, Gevorgyan2021}. Shorter composite pulses are also available from the \textsc{scrofulous} family \cite{Cummins2003}, but these are less effective at suppressing errors, and with the exception of \NOT\ gates require some unusual rotation angles for individual sub-pulses.

Tackling off-resonance errors is also difficult for pulse flip angles other than $180^\circ$. The \textsc{corpse} and short-\textsc{corpse} sequences  \cite{Cummins2003} give moderate error suppression, but again require unusual rotation angles. More recently these pulses have been placed in a wider context \cite{Kukita2022b}, but the original solutions remain among the most promising. Of more interest are the ConCatenated Composite Pulses (CCCPs) \cite{Ota2009, Ichikawa2012, Bando2013}, which provide simultaneous compensation of off-resonance and pulse-strength errors for arbitrary flip angles, and which have been demonstrated in NMR experiments \cite{Bando2020}.

Finally there has been significant theoretical interest in exploring the limits of error suppression with composite pulses, beyond the specific iterative approach to suppression of $B_1$ errors in \NOT\ gates \cite{Jones2013a}. Note that the interest within QIP is usually in obtaining very precise quantum gates in the presence of moderate underlying errors, the opposite of the situation in conventional NMR which usually seeks moderate performance over very wide ranges of parameter values. A key result is that there is no limit in principle to the accuracy that can be achieved as existing pulse designs can always be improved using methods similar to those used to derive the Solovay--Kitaev theorem \cite{Brown2004, Brown2005}. The original paper is a challenging read, but a more detailed explanation in more conventional NMR notation is available \cite{Alway2007}, which also clarifies the need for sufficiently accurate inverse pulses when using the Solovay--Kitaev construction. This is not a problem for  $B_1$ errors, as a $\theta_{-x}$ pulse remains an accurate inverse for a $\theta_x$ pulse, but care is needed when seeking to correct off-resonance errors, as in this case the errors will add up in the sequence $\theta_x\theta_{-x}$ instead of cancelling out \cite{Alway2007}. There are also specific results available for the case of pulse strength errors \cite{Low2014}, where it is possible to draw analogies between composite pulses and filter designs \cite{Low2016a}.

\subsection{Choosing an approach}
Shaped pulses developed for applications in conventional NMR have frequently used restricted forms. This choice seems to have been driven firstly by a desire for pulses which either vary smoothly or which change sharply at only a small number of points, thus imposing fewer demands on the implementation hardware, and secondly by a belief that the number of controllable parameters should be kept small to reduce the computation time required. Both of these concerns are now unwarranted, due to the design of modern spectrometers with direct digital synthesis, which can produce even complicated waveforms with comparative ease, and the rapid progress in computer power, traditionally summarised in Moore's laws \cite{Moore2006}. With computing power increasing by an order of magnitude every five years \cite{Roser2023}, problems that were very challenging thirty years ago are now straightforward.

These concerns also led to a concentration on algorithms that avoid gradients. Superficially it appears that fidelity gradients can only be calculated using finite difference methods, and this requires $n+1$ function evaluations for a function with $n$ input parameters. If these inputs are simply digitised amplitudes, then there will be $n$ sub-propagators to calculate for each function evaluation, leading to an apparent $O(n^2)$ time complexity for gradient-based methods, compared to $O(n)$ for methods that only use function values directly. Gradient-free methods also permit solutions to the possibility of local minima, as described in Section~\ref{sec:OA} above. This approach has been explored within QIP as the chopped random basis (CRAB) \cite{Caneva2011} and related algorithms \cite{Mueller2022}. However, a key result about optimal control landscapes is that the great majority of control problems are in fact free of such traps \cite{Rabitz2004, Rabitz2006, Chakrabarti2007, Wu2008}, suggesting that such concerns are in fact unlikely to be important.

Avoiding gradients is also usually unnecessary as there are methods to find gradients more efficiently by storing partial results, an example of a time--memory tradeoff \cite{Hellman1980}. Within NMR this is usually implemented through the gradient ascent pulse engineering (GRAPE) algorithm \cite{Khaneja2005}, which has applications in both NMR QIP and more conventional NMR studies, and which is explored in detail in Section~\ref{sec:GRAPE}. Other implementations of QIP have largely concentrated on the earlier Krotov family of algorithms \cite{Krotov1995,Zhu1998, Maday2003, Maximov2008}. The principal difference between these approaches is that the GRAPE family uses gradient calculations to update all the points in a pulse shape simultaneously, while the Krotov family sweeps forwards and backwards across the shape. Although these two families superficially appear quite different, it is possible to describe them, and possible hybrids, within a unified framework \cite{Machnes2011, Schulte-Herbruggen2012}. Gradient techniques can also be applied within the CRAB family, giving rise to the gradient optimization of analytic controls (GOAT) scheme \cite{Machnes2018}.

\section{GRAPE} \label{sec:GRAPE}
Gradient ascent pulse engineering (GRAPE) \cite{Khaneja2005} can refer to a wide range of related algorithms for optimal control, usually but not always within the context of NMR. Implementations can differ in the choice of underlying fidelity function, the presence of penalty functions, and the choice of optimization algorithm, but are all united by a common approach to the calculation and use of fidelity gradients.
\begin{figure}
\begin{center}
\includegraphics{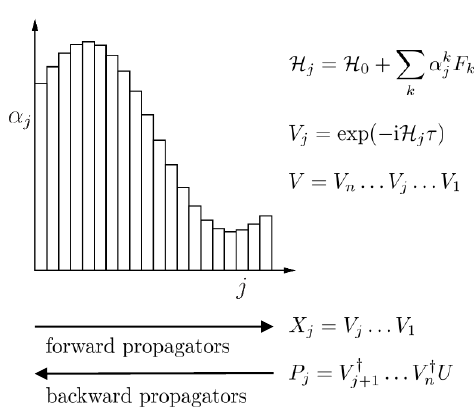}
\end{center}
\caption{The GRAPE trick allows the inner product \inner{U}{V} to be rewritten in terms of forward and backward propagators as \inner{P_j}{X_j}, which enables gradients to be calculated efficiently by storing intermediate values.}\label{fig:GRAPE}
\end{figure}

Here I concentrate on applications within NMR QIP, and so I largely consider the standard unitary fidelity, equation~\ref{eq:Ufid}. The Hamiltonian is normally assumed to be piecewise continuous (although more general forms have also been considered \cite{Rasulov2023}), where the $j$th Hamiltonian is applied for a fixed time $\tau$, and takes the form of a sum over the drift Hamiltonian and all possible control Hamiltonians scaled by their amplitudes,
\begin{equation}
{\mathcal H}_j={\mathcal H}_0+\sum_k\alpha^k_j F_k,
\end{equation}
as shown in Figure~\ref{fig:GRAPE}. Here the sum over $k$ can run over $x$ and $y$ (to allow phase control as well as amplitude control) and also over multiple nuclear species in a heteronuclear system. As usual the overall propagator is given by the time ordered product
\begin{equation}
V=V_n\dots V_j\dots V_1\label{eq:Vprod2}
\end{equation}
with sub-propagators
\begin{equation}
V_j=\exp(-\rmi{\mathcal H}_j\tau).\label{eq:Vjtau}
\end{equation}
This restriction to fixed equal time intervals is not essential to what follows, but is a common and convenient approach, reflecting the way shaped pulses are encoded within NMR hardware.

The original authors considered a wide range of fidelity functions \cite{Khaneja2005}, corresponding to different tasks and to the presence of different assumptions about relaxation, but for optimising unitary transformations they seek to maximise
\begin{equation}
\Phi_4=|\inner{U}{V}|^2=\inner{U}{V}\inner{V}{U}\label{eq:Phi4}
\end{equation}
where the inner product between two operators is defined as
\begin{equation}
\inner{U}{V}={\rm tr}(U^\dag V).
\end{equation}
This trace form for an inner product may appear unfamiliar, but is in fact precisely how the inner product between two kets is defined if the kets are written explicitly as matrices: the product of a complex conjugated row matrix (representing a bra) by a column matrix (representing a ket) gives a one-by-one matrix, and taking the trace of this matrix converts the single element to a scalar as desired. Note that $\Phi_4$ differs from the conventional unitary fidelity, equation~\ref{eq:Ufid}, by a normalisation factor. For a system of $q$ qubits $U^\dag U$ is the identity matrix of size $2^q$, and so
\begin{equation}
{\mathcal F}=\Phi_4/4^q,
\end{equation}
but if one is seeking to maximise the fidelity the precise normalisation is irrelevant as long as one is consistent.

The next stage is to rewrite the inner product in an equivalent form
\begin{equation}
\begin{split}
\inner{U}{V}
 & = {\rm tr}(U^\dag V_n\dots V_{j+1}V_j\dots V_1) \\
 & = {\rm tr}([V_{j+1}^\dag\dots V_n^\dag U]^\dag[V_j \dots V_1]) \\
 & = {\rm tr}(P_j^\dag X_j) \\
 & = \inner{P_j}{X_j}
\end{split}
\end{equation}
where the second line uses the standard identity $ABC=(C^\dag B^\dag A^\dag)^\dag$ and the fact that the adjoint is self inverse. Here
\begin{equation}
X_j=V_j \dots V_1
\end{equation}
is the \textit{forward propagated} operator up to the $j$th time period, and
\begin{equation}
P_j=V_{j+1}^\dag\dots V_n^\dag U
\end{equation}
is the \textit{backward propagated} target. In this notation
\begin{equation}
\Phi_4=\inner{P_j}{X_j}\inner{X_j}{P_j}\label{eq:Phi4PX}
\end{equation}
for \textit{any} value of $j$, with the conventional form, equation~\ref{eq:Phi4}, corresponding to the choice $j=n$.

This form is far more convenient for calculating derivatives,
\begin{equation}
\frac{\partial\Phi_4}{\partial\alpha^k_j}
=2{\rm Re}\left(\inner{P_j}{\frac{\partial X_j}{\partial\alpha^k_j}}\inner{X_j}{P_j}\right),
\end{equation}
which follows from the product rule, the linearity of the trace function, the fact that $P_j$ is independent of $\alpha^k_j$, and equation~\ref{eq:dy2}. The forward propagator $X_j$ does depend on the $j$th set of control amplitudes, but only through the final sub-propagator,
\begin{equation}
\frac{\partial X_j}{\partial\alpha^k_j}=\frac{\partial V_j}{\partial\alpha^k_j}V_{j-1}\dots V_1.
\end{equation}
Up to this point everything is \textit{exact}.

Calculating the derivative of the sub-propagator is more challenging, but if the control amplitudes are sampled finely then $\tau$ will be small enough that a linear approximation can be used,
\begin{equation}
\frac{\partial V_j}{\partial\alpha^k_j}\approx-\rmi \tau F_k V_j,\label{eq:dVbyda}
\end{equation}
as discussed below. Putting this all together leads to the key result
\begin{equation}
\frac{\partial\Phi_4}{\partial\alpha^k_j}
\approx-2{\rm Re}\left(\inner{P_j}{\rmi\tau F_kX_j}\inner{X_j}{P_j}\right),\label{eq:dPhi4byda}
\end{equation}
which is accurate to first order in $\tau$ \cite{Khaneja2005}. The significance of this form is that the forward and backward propagators can be calculated efficiently if partial results are stored. Since $X_j=V_jX_{j-1}$, and so on, it is only necessary to calculate each sub-propagator once, and then to multiply everything out twice: forwards to obtain the $X$ matrices and backwards to obtain the $P$ matrices. This permits gradients to be estimated is a time $O(n)$, that is linear in the number of control points rather than the quadratic dependence observed for naive finite difference methods.

Similar formulae can be derived for a range of alternative fidelity measures, and including the effects of non-unitary evolution. While these methods have important applications in conventional NMR \cite{Khaneja2005a, Kehlet2007, Coote2017}, they are rarely relevant to NMR QIP and will be largely ignored here. Writing an implementation of GRAPE is fairly straightforward using a high-level computing language which provides optimization routines. Alternatively, implementations are available as packages written in Matlab (Dynamo \cite{Machnes2011}, Spinach \cite{Hogben2011, Goodwin2016}), Python (QuTiP \cite{Johansson2012, Johansson2013}), Julia \cite{Goerz2022}, and C (SIMPSON \cite{Tosner2009}).

\subsection{Approximate derivatives}
In the section above I simply asserted that equation~\ref{eq:dVbyda} provides an approximate formula for the derivative of a sub-propagator with respect to one of the control amplitudes. Before turning to the correct formula for the exact derivative it is useful to consider a simple justification for this form, which also shows why it is only approximate and indicates the conditions under which the approximation is a good one. Start by writing
\begin{equation}
\frac{\partial V_j}{\partial\alpha^k_j}=\lim_{\delta\rightarrow0}\frac{\exp(-\rmi[{\mathcal H}_j+\delta F_k]\tau)-\exp(-\rmi{\mathcal H}_j\tau)}{\delta}
\end{equation}
and note that the fundamental problem in evaluating this is that $F_k$ will not normally commute with $\mathcal H_j$, which makes evaluation of the first matrix exponential complicated. However, as $\delta$ is small and small evolutions almost commute with everything, this can be approximated as
\begin{equation}
\begin{split}
\exp(-\rmi[{\mathcal H}_j+\delta F_k]\tau)
&\approx\exp(-\rmi{\mathcal H}_j\tau)\exp(-\rmi\delta F_k\tau)\\\label{eq:HplusDelta}
&\approx\exp(-\rmi\delta F_k\tau)\exp(-\rmi{\mathcal H}_j\tau).
\end{split}
\end{equation}
Choosing the second form, and using the fact that as $\delta$ is small a series expansion can be used for the first exponential term, giving
\begin{equation}
\begin{split}
\frac{\partial V_j}{\partial\alpha^k_j}
&\approx\lim_{\delta\rightarrow0}\left(\frac{{\mathbf{1}}-\rmi\delta F_k\tau+O(\delta^2)-{\mathbf{1}}}{\delta}\right) 
\exp(-\rmi{\mathcal H}_j\tau)\\
&\approx -\rmi\tau F_k V_j
\end{split}
\end{equation}
as stated previously. The flaw in this argument can be seen by instead choosing the first approximate form in equation~\ref{eq:HplusDelta}, which leads to
\begin{equation}
\frac{\partial V_j}{\partial\alpha^k_j}\approx -\rmi\tau V_j F_k,
\end{equation}
and since $V_j$ and $F_k$ will not normally commute these two forms will be different, and neither of them will be correct. The solution to this is simply to note that if $\tau$ is small enough then $V_j$ will be a small evolution that almost commutes with everything, and so the two forms are almost the same and are both approximately correct, with equation~\ref{eq:dVbyda} chosen for convenience in subsequent calculations. As stated in \cite{Khaneja2005} this result is only valid to first order in $\tau$. For this reason, the standard approximate gradient, equation~\ref{eq:dPhi4byda}, becomes more accurate as the shape of the pulse is sampled more finely. Fortunately the linear time scaling achieved by GRAPE means that fine enough division is normally practical.

\subsection{Exact derivatives}
While it is possible to use these approximate derivatives, it would be desirable to find a more precise formula \cite{Motzoi2009}, as this will give much better convergence with more sophisticated optimization algorithms such as BFGS \cite{DeFouquieres2011, Machnes2011, Jensen2021}. The route to an exact formula has been known for some time \cite{Aizu1963, Wilcox1967}, and has been applied within conventional NMR \cite{Levante1996}. The exact derivative of the exponential of a sum of two non-commuting operators $A$ and $xB$ with respect to $x$ at $x=0$ can be evaluated in the eigenbasis of $A$ as
\begin{multline}
\left\langle \xi_l \left| \frac{\partial}{\partial x} \rme^{A+xB} \right|\xi_m \right\rangle \\=
\begin{cases}
\matrixelem{\xi_l}{B}{\xi_m}\rme^{\xi_l},   & \mathrm{if}~\xi_l = \xi_m,\\
\matrixelem{\xi_l}{B}{\xi_m} \frac{\rme^{\xi_l}-\rme^{\xi_m}}{\xi_l - \xi_m}, & \text{otherwise},
\end{cases} \label{ch2:exactderivative}
\end{multline}
where $A\ket{\xi_l} = \xi_l\ket{\xi_l}$. This result is derived in Appendix A of \cite{Machnes2011}. This approach requires $\mathcal H_j$ to be diagonalized at each point, but the resulting eigenvectors and eigenvalues can be reused to calculate matrix exponentials, replacing the more normal combination of the scaling and squaring and Pad\'e approximant methods \cite{Moler2003}.

\subsection{Approximate evaluation of propagators}
The section above describes how to perform optimizations more accurately, but there remains some value in methods for performing approximate calculations as rapidly as possible. This is principally useful to obtain good initial guesses for a control pulse which can then be optimized by more precise methods. One approach which initially appeared promising was the method of Gradient Ascent Without Matrix Exponentiation (GRAWME) \cite{Bhole2018}, which replaces all the matrix exponentials in a calculation by approximate forms. This method has been superseded by the realisation that phase-only control, discussed in the next section, gives even greater speed gains while retaining full accuracy, but the idea remains of historical interest, and similar ideas have been applied in other contexts \cite{Peterson2020}.

GRAWME begins by writing the control fields in terms of a time-varying amplitude and phase, rather than the $x$ and $y$ amplitudes, to get
\begin{equation}
{\mathcal H}_j={\mathcal H}_0+A_j\left(\cos\phi_j\, F_x +\sin\phi_j\, F_y\right).\label{eq:HAphi}
\end{equation}
This allows $V_j$ to be rewritten as
\begin{equation}
V_j=\rme^{-\rmi\phi_jF_z}\,V_j^x\,\rme^{\rmi\phi_jF_z}\label{eq:Vjx}
\end{equation}
where
\begin{equation}
V_j^x=\exp(-\rmi[{\mathcal H_0}+A_jF_x]\tau)
\end{equation}
is the equivalent operator with all the amplitude along $x$. This operator is the sum of two non-commuting observables, and so requires explicit matrix exponentiation \cite{Moler2003}. It can, however, be approximated using the Trotter--Suzuki form \cite{Suzuki1976, Suzuki1985, Childs2021}
\begin{equation}
V_j^x\approx\rme^{-\rmi{\mathcal H_0}\tau/2}\,\rme^{-\rmi A_jF_x\tau}\,\rme^{-\rmi{\mathcal H_0}\tau/2}
\end{equation}
which is accurate to third order in $\tau$. Here the first and third terms are fixed; the central term depends on $A_j$ but can be easily evaluated as the eigenbasis is fixed, and so can be made diagonal with a known fixed basis transformation which interconverts $F_x$ and $F_z$. Putting everything together gives
\begin{equation}
V_j^x\approx\rme^{-\rmi\phi_jF_z}\,W_1\,\rme^{-\rmi A_jF_z}\,W_2\,\rme^{\rmi\phi_jF_z}
\end{equation}
with all explicit matrix exponentials now diagonal in the computational basis. The two basis transformations are defined by
\begin{equation}
W_1=\rme^{-\rmi{\mathcal H_0}\tau/2}\,{\mathrm H}^{(q)},\quad
W_2={\mathrm H}^{(q)}\,\rme^{-\rmi{\mathcal H_0}\tau/2},
\end{equation}
where ${\mathrm{H}}^{(q)}$ is the $q$-qubit Hadamard gate, which converts between the $x$ and $z$ basis. As these are independent of $A_j$ and $\phi_j$ they need only be calculated once. For the situations typical in the design of NMR GRAPE pulses the fractional error in the evaluation of fidelities is around $10^{-6}$, which is negligible in many cases \cite{Bhole2018}.

Avoiding explicit matrix exponentials (or more precisely only evaluating matrix exponentials in a diagonal basis, where the calculation is easy) will clearly speed up the evaluation of propagators, but unfortunately the overall gain is only by a constant factor. The most time-consuming step is now matrix multiplication, and like matrix exponentiation this is an $O(N^3)$ process, where $N=2^q$ is the dimension of the vector space. Further constant gains can be obtained by careful coding of multiplications involving diagonal matrices \cite{Bhole2018}, but it is not possible to entirely avoid full matrix multiplications, and the overall speed gains observed were around a factor of 10. Extensions to higher order approximations have also been explored \cite{Yang2022}.

\subsection{Phase-only control}\label{sec:phaseonly}
Phase-only control \cite{Skinner2006} has several significant advantages over general control for the design of GRAPE pulses. The derivations above assume that the amplitudes of the $x$ and $y$ components of the control fields are varied independently, or equivalently that the amplitude and phase of the RF field are both control variables. In phase-only control the amplitude is held at some fixed value $A$ and only $\phi_j$ is allowed to vary. This means that $V_j^x$ in equation~\ref{eq:Vjx} is constant, and only has to be calculated once, so there is no reason to use approximations.

As before the phase shift operators are diagonal, and so easy to calculate. The exact derivative is also easy to calculate directly as
\begin{equation}
\frac{\partial V_j}{\partial\phi_j}=-\rmi F_z V_j+\rmi V_j F_z=\rmi\left[V_j,F_z\right].
\end{equation}
If desired this approach can be extended to calculate the exact Hessian directly \cite{Violaris2021}, rather than approximating it by BFGS methods. For maximum efficiency it is important to use the diagonal structure of the phase shift operators to perform the relevant multiplications rapidly, rather than naively using a full matrix form \cite{Bhole2018}.

Phase-only control has the further significant advantage of removing any need to apply penalty functions to discourage excessive RF amplitudes, as the amplitude is simply fixed at some desired value. This will also remove any transient errors arising from amplitude changes, except at the start and end of the pulse. If a smoothly varying amplitude is desired instead, then it is easy to modify the calculation to use a pre-determined value for $A_j$ at each point. Experience suggests that phase-only control is in practice almost as flexible as full control as long as the time step $\tau$ is chosen small enough, and the efficiency of the calculations more than makes up for any increase in the number of control parameters. Note that phase only control takes shaped pulse design back to its origin in composite pulses, and phase-only shaped pulses can be interpreted as very long composite pulses \cite{Warren1988}. An important example from conventional NMR is the use of binomial solvent-suppression sequences \cite{Hore1983}, although these only use phases of $0$ and $180^\circ$.

\subsection{Subsystem control}
The methods above can provide significant speed-ups, making GRAPE pulses an entirely practical method for implementing quantum logic gates in systems with three or four spins, but the fundamental scaling of the computational time required with the size of the spin system remains a problem. As noted above, the time required for elementary matrix multiplications scales as $O(N^3)$, where $N=2^q$ is the dimension of the Hilbert space for a system of $q$ qubits. As a consequence the time required to design a GRAPE pulse increases by a factor of at least 8 for every additional spin in the system, and in practice the growth is often worse as more selective control usually requires a longer control sequence.

A partial solutions to this is provided by subsystem control \cite{Ryan2008}. Suppose that one wishes to design a single-qubit gate in the four-qubit system provided by the \nuc{13}{C} nuclei in labelled crotonic acid (Figure~\ref{fig:4qubitcrot}). These four spins form a rough linear chain, with large couplings (over 40\,Hz) between nearest neighbours and smaller long range couplings (under 10\,Hz). This system can be fairly well modelled as a pair of three-spin systems, one made up from the first three spins and the other from the last three, with the omitted spin and all couplings to it simply dropped from the two subsystem Hamiltonians. A control sequence which performs a gate on the four-spin system should also perform an equivalent action on the two three-spin subsystems fairly well, and \textit{vice versa}. The fidelity of the operation in a four-qubit system can be approximated as the average fidelity over the two three-qubit subsystems, and it is considerably faster to perform calculations with two three-qubit systems than with a single four-qubit system. The equivalence of the fidelities will not be perfect, but it is easy to check the fidelity of the subsystem solution for the full Hamiltonian, and if necessary to complete the optimization over the full Hamiltonian starting from the subsystem solution as a good initial guess.

The subsystem approach can be taken further, describing crotonic acid as a combination of three different two-qubit subsystems, retaining only the nearest neighbour pairs with large couplings. For a controlled gate it is clearly essential to include at least those couplings directly involved in the control process, but if the aim is to design a single-qubit gate then the most extreme simplification, modelling the system as four independent single-qubit subsystems, can be a useful start.

While subsystem control has proved useful even in small systems, its real power comes into play in much larger spin systems. For example, the 12 qubit system implemented with seven \nuc{13}{C} and five \nuc{1}{H} spins \cite{Lu2017a} is too large to simulate directly, and was instead simulated using either two non-overlapping subsystems of six spins each \cite{Li2019a}, which does not allow full control, or five overlapping subsystems with between two and four spins in each \cite{Peterson2020}, which enables every pair of spins to be accessed either directly or indirectly.

Similar approaches have been used to simulate the quantum circuits used in NISQ (noisy intermediate-scale quantum) devices \cite{Preskill2018, Arute2019etal}. The presence of decoherence in such systems means that only an approximate simulation is required, permitting the effective simulation of circuits previously claimed to lie beyond the limits of simulation \cite{Ayral2023}. Such approaches cannot, however, be used to simulate error-free quantum systems, raising concerns as to whether subsystem control can be used effectively in true quantum computers.

\subsection{Single-spin control}\label{sec:SSC}
A special case occurs when optimal control is performed on an ensemble of single-qubit systems, either as an extreme example of subsystem control or for applications in conventional NMR such as the design of broadband pulses or pulses that selectively excite particular frequency bands \cite{Kobzar2004, Kobzar2005, Gershenzon2008, Kobzar2012, Nimbalkar2013, Coote2018, Behera2020, Coote2021, Haller2022, Slad2022, Joseph2023}. Such pulses can, of course, be designed using any of the methods described above, but for single spin control significant speed-ups are possible by taking advantage of the small size of the vector space. In particular the sub-propagator $V_j$ and its derivatives can be easily evaluated analytically, rather than using the numerical methods which are required for larger spin systems.

For single-spin control state-to-state fidelity measures are particularly interesting, as the relevant state space is small. In conventional NMR it is common to seek pulses that perform correctly for a spin initially along the $z$ axis of the Bloch sphere, such as inversion or excitation pulses. This can be achieved within QIP by using the fidelity for the initial state \ket{0}, and the approach can obviously be generalised to optimise the performance for any particular starting state.

Another common problem in conventional NMR is refocusing pulses, which perform well for spins in the $xy$ plane. These could be found by optimising over two orthogonal states in the $xy$ plane, such as $\ket+=(\ket0+\ket1)/\sqrt2$, which lies along the $x$-axis, and $\ket R=(\ket0+\rmi\ket1)/\sqrt2$, which lies along $y$. However, any single spin operation which performs correctly along two orthogonal axes will also perform correctly along the third, and there is no substantive difference between optimising the state-to-state fidelity averaged over any two orthogonal states and optimizing the unitary fidelity.

This is most easily seen by considering the form of $U^\dag V$, which describes any erroneous transformation that $V$ performs in addition to the desired transformation $U$. For a single spin this corresponds to some rotation around some axis on the Bloch sphere, and any such rotation can only leave two particular states unaffected, these being the states lying along the rotation axis for $U^\dag V$. The sole exception to this general rule is the identity operation, which leaves the entire Bloch sphere unaffected. Thus if $V$ performs $U$ precisely for any two states which are not on opposite sides of the Bloch sphere then $U^\dag V$ must be the identity operation, and so $V$ must be equal to $U$.

Thus the only high-fidelity single spin controls worth considering are unitary controls and controls for a single initial state: refocusing pulses are simply equivalent to unitary rotations. Obviously any unitary control can be used as a good state-to-state transfer, but a good state-to-state pulse can be significantly shorter or more robust. More interestingly, the process can be partially reversed: there is a simple procedure to convert some state-to-state pulses into unitary pulses with twice the rotation angle and taking twice as long \cite{Luy2005}.

For single-spin control it can also be useful to code parts of the algorithm directly by hand \cite{Buchanan2024} rather than using standard libraries, particularly when using interpreted high-level languages such as Matlab. Such languages have highly optimized routines for operations such as matrix exponentiation, which are particularly effective with large matrices, but when using two-by-two matrices to describe single spins the overhead imposed by calling routines and using standard data structures can far outweigh the relatively small amount of time spent actually calculating results. With a two-by-two matrix it is perfectly possible to simply store the four elements as individual values, and to multiply matrices by hand, explicitly coding the result for each element. While such code can be difficult to maintain, the resulting speed gains can be very significant. The gains are smaller for compiled languages, but still worthwhile.

For unitary propagators corresponding to traceless Hamiltonians an even more compact approach is possible: all such propagators have the form
\begin{equation}
U=\begin{pmatrix}\alpha&\beta\\-\beta^*&\alpha^*\end{pmatrix}, \quad |\alpha|^2+|\beta|^2=1,
\end{equation}
and so it is only necessary to evaluate two elements and the whole matrix is known. Similarly, the trace of such a matrix is twice the real part of either diagonal element, and so matrix traces can be evaluated efficiently \cite{Buchanan2024}. These observations are closely related to the use of quaternions to describe single spin rotations \cite{Cummins2003, Emsley1992a}.

\subsection{Decoupling passive spins}\label{sec:passive}
As discussed in section~\ref{sec:spinsystem}, it is quite common to implement a QIP protocol using a spin system containing more spin-\half\ nuclei than the number of qubits required. In particular the four \nuc{13}{C} nuclei in labelled crotonic acid provide an extremely popular four qubit system, but these four spins are embedded in a larger system containing two distinguishable \nuc{1}{H}{ nuclei, providing possible qubits, and three \nuc{1}{H} nuclei in a methyl group which could be used as a further qubit. The system has been used to implement seven qubit experiments \cite{Knill2000}, but the most common approach is to reduce the spin system to four qubits by decoupling the \nuc{1}{H} nuclei \cite{Boulant2002}, using conventional broadband decoupling sequences \cite{Ryan2005} such as WALTZ-16 \cite{Shaka1983b}.

While this idea seems obvious, it works less well than one might hope, and many experiments which use labelled crotonic acid as a four qubit system suffer from very significant signal losses, which are rarely explicitly acknowledged and even more rarely explained. The explanation is that while broadband \nuc{1}{H} decoupling is very effective at removing the heteronuclear couplings during free evolution, it is far less effective in the presence of simultaneous \nuc{13}{C} irradiation, as happens during GRAPE pulses, due to uncontrolled Hartmann--Hahn transfers \cite{Levitt1991}. It is straightforward to perform a brute force simulation of the evolution under the full nine spin Hamiltonian in the presence of a decoupling sequence with realistic RF power, and when this is done the apparently mysterious signal losses are replicated \cite{Bhole2020}.

One possible solution to this is to remove the heteronuclear couplings by spin echoes rather than continuous decoupling \cite{Boulant2003}, but this only works where controlled gates are constructed from sequences of short pulses and longer delays, rather than being implemented directly as long GRAPE sequences. It would be desirable to find some way in which the couplings to these \textit{passive} spins, which play no role in the controlled spin system of \textit{active} spins, but are simply coupled to it, could be ignored without the need to explicitly decouple them.

This could be achieved by preparing the passive spins in a pure state, or equivalently as part of a pseudo-pure state, in effect selecting a subset of the components in the multiplet. If all the \nuc{1}{H} spins are in state \ket{0} then the effect of the heteronuclear couplings is to cause a shift, rather than a splitting, and they can simply be absorbed into the chemical shift, while the homonuclear \nuc{13}{C} couplings remain as normal. This method has been used to implement a five qubit system in crotonic acid \cite{Ryan2005a} by using the four \nuc{13}{C} nuclei and the spin-\half\ component of the methyl group, while setting the remaining \nuc{1}{H} nuclei to state \ket{0}. With this approach it is essential that the passive spins remain in \ket{0}, much as in TROSY experiments \cite{Pervushin1997}, and so it is vital to avoid accidental excitation by RF fields.

An even simpler approach to this problem is to leave the passive spins in a highly mixed state, and then design pulses which are insensitive to the heteronuclear couplings. As the passive \nuc{1}{H} spins remain in a fixed state during a \nuc{13}{C} pulse sequence, their effect is simply to apply a frequency offset which depends on their state and so is different for different molecules in the ensemble. The system of nine spins can be treated as 32 different subsystems, corresponding to the 32 possible states of the \nuc{1}{H} nuclei, with a subtly different four spin Hamiltonian for each subsystem. The fidelity of a pulse sequence can then be averaged over these subsystems, and the resulting GRAPE pulse will correctly address the active spins whatever states the passive spins happen to be in \cite{Bhole2020}. As the three methyl protons are indistinguishable this can be achieved more efficiently by using a weighted average over the 16 distinguishable \nuc{1}{H} spin states. Broadband decoupling should be applied to \nuc{1}{H} during acquisition, to simplify the observed spectra, but \textit{must not} be applied during the logic gates, to ensure that the passive spins remain passive.

\section{Pseudo-pure states}
\label{sec:PPS}
\begin{figure*}
\begin{center}
\includegraphics[width=16cm]{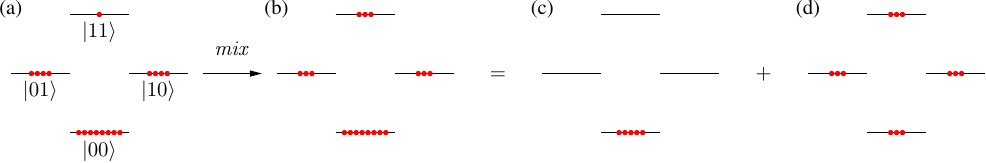}
\end{center}
\caption{Preparing a pseudo-pure \ket{00} state in a homonuclear two qubit system. A thermal state (a) has higher populations in the lower levels, shown exaggerated here. A mixing process is applied to equalise populations in the upper levels, leaving the lowest level untouched (b). The result can be treated as a mixture of the desired pure state (c) and the maximally mixed state (d) with equal populations in every level.}\label{fig:PPS}
\end{figure*}
Quantum information protocols use unitary transformations to achieve tasks which are impossible for purely classical devices, but to obtain the correct results it is essential that the system starts in a well-defined initial state, usually taken as the state \ket{00\dots0}, with all qubits in state \ket0. As this initial state must be prepared, whatever the state of the system before the initialisation step, the initialisation process is obviously non-unitary, and in particular must be a process, such as cooling, which is capable of taking the system from a mixed state to a pure state.

Unfortunately the non-unitary processes available within conventional NMR are not capable of achieving this. Evolution under the drift Hamiltonian or control Hamiltonians is unitary, while decoherence ($T_2$ relaxation) takes the system to a \textit{more mixed} state. The same is true of processes such as gradient dephasing and phase cycling, which can be thought of as controllable decoherence. The sole exception is $T_1$ relaxation to the thermal state, but while this can increase the purity of the spin state it remains very highly mixed.

The standard solution within NMR QIP is to prepare a \textit{pseudo-pure state}, also called an \textit{effective pure state}, as shown in Figure~\ref{fig:PPS} for a two qubit system. The underlying idea is to equalise the populations of all the excited states, leaving the ground state, which has the highest population at thermal equilibrium, untouched. The resulting mixed state can be reinterpreted as a mixture of the desired pure state and the maximally mixed state. Since the maximally mixed state does not evolve under unitary transformations, and gives no detectable NMR signal, this pseudo-pure state behaves just like a genuine pure state except that the signal is scaled down, reflecting the effective purity.

It is important to remember that a mixed state has no unique decomposition, and the belief that a pseudo-pure state really is a mixture of the pure state and the maximally mixed state is an example of the \textit{preferred ensemble fallacy} or \textit{partition ensemble fallacy} \cite{Kok2000}. For this reason it is generally not possible to use NMR methods to perform tests of quantum mechanics, as the results can usually be reinterpreted using a different decomposition \cite{Braunstein1999}. However it remains true that apart from a scaling factor NMR experiments on pseudo-pure states give precisely the same results as experiments on pure states, as demonstrated by pure state NMR implementations of Deutsch's algorithm \cite{Anwar2004b} and Grover's algorithm \cite{Anwar2004a}, which are indistinguishable from their pseudo-pure counterparts. Furthermore, attempts to describe NMR QIP experiments in purely classical terms  \cite{Schack1999} appear to be impossible.

\subsection{Single spins}
The case of a single isolated spin-\half\ nucleus is special, as no preparation sequence is necessary. The thermal state can be written in NMR notation as
\begin{equation}\label{eq:rhoppIz}
\rho=\half E+p\,I_z,
\end{equation}
with the polarization $p\approx\hbar\omega/2k_BT\sim10^{-5}$. Here $I_z$ is a deviation density matrix, with trace equal to zero, rather than a proper density matrix, with trace equal to one. This can be rewritten as
\begin{equation}\label{eq:rhopp0}
\rho=(1-p)\mathbf{1}/2+p\ket0\bra0
\end{equation}
where $\mathbf{1}/2=\half E$ is the maximally mixed state for a single spin and $\ket0\bra0$ is a proper density matrix corresponding to the pure state \ket0, and so this is already a pseudo-pure state, as discussed in Section~\ref{sec:states}.

This is why the Bloch sphere picture can be directly transferred to describe single spin NMR, ultimately leading to the success of the vector model \cite{Bloch1956, FreemanSCbook}. The conventional NMR approach is built around traceless observables, as done in equation~\ref{eq:rhoppIz}, dropping the undetectable $\half E$ term. The polarization term $p$ could be retained, but as this simply scales the size of the NMR signal, and the absolute signal size has no fundamental meaning, it is convenient to rescale everything such that $p=1$. This is not true for larger spin systems, where pseudo-pure states are quite different from thermal states, and intuitions from conventional NMR are far less applicable to QIP systems.

\subsection{Two spins}
For two spins the thermal state can be written in NMR notation as $I_z+S_z$, but this is no longer a pseudo-pure state. The desired state is now
\begin{equation}
\rho=(1-p)\mathbf{1}/4+p\ket{00}\bra{00},
\end{equation}
with
\begin{equation}\label{eq:rhopp00}
\ket{00}\bra{00}=\half\left(\half E+I_z+S_z+2I_zS_z\right),
\end{equation}
and other initial pseudo-pure states can be written in a similar way as
\begin{equation}\label{eq:rhoppab}
\begin{split}
\ket{01}\bra{01}&=\half\left(\half E+I_z-S_z-2I_zS_z\right),\\
\ket{10}\bra{10}&=\half\left(\half E-I_z+S_z-2I_zS_z\right),\\
\ket{11}\bra{11}&=\half\left(\half E-I_z-S_z+2I_zS_z\right).
\end{split}
\end{equation}
To generate a pseudo-pure state it is necessary to make an appropriate mixture of the three population states, including the two-spin order population term. Note that it is obviously possible to include the $\half E$ component in with the maximally mixed part, and so it is not necessary to specifically generate this.

Reversing this argument, single spin polarization terms such as $I_z$ do not correspond to pure states, but must represent mixed states. This is entirely unsurprising, as terms like $I_z$ indicate that spin $S$ is in a completely mixed state. It is, however, easy to prepare states corresponding to a single pure qubit, with the remaining qubits in maximally mixed states, which are used in the DQC1 model of computation \cite{Knill1998a}.

\subsection{Preparation methods}\label{sec:simplePPS}
Just like for pure states, the preparation process for pseudo-pure states must be non-unitary, except for single spin systems where no preparation is required. The easiest way to see this is to note that the eigenvalues of the density matrices are different for pseudo-pure and thermal states, and so these cannot be related by a unitary transformation, which always leaves the eigenvalues unchanged. As both states are diagonal in the computational basis, these eigenvalues can simply be read off directly as the state populations. In a pseudo-pure state for a two-spin system, three states will have the same population, while the state corresponding to the desired pure state will have a higher population. By contrast the populations in the thermal state will be more diverse, with three distinct values in a homonuclear two-spin system and four distinct values in the heteronuclear case.

Methods for preparing pseudo-pure states can be divided into three broad categories. The conceptually simplest approach is \textit{logical labelling}, which simply uses a subset of levels within a larger spin system which happen to have the right pattern of populations \cite{Gershenfeld1997, Chuang1998b}. For example, a two qubit computer can be encoded using three physical spin-\half\ nuclei by assigning physical state $\ket{\alpha\alpha\alpha}=\ket{000}$ to logical \ket{00} and physical states \ket{\beta\beta\alpha}, \ket{\beta\alpha\beta} and \ket{\alpha\beta\beta} to logical \ket{01}, \ket{10} and \ket{11} in some order. It is obviously necessary to use a larger number of physical spins than logical qubits, but the overhead is not too large \cite{Gershenfeld1997}.

The simplicity of the preparation sequence comes at a cost in the complexity of implementing quantum gates, as even single-qubit gates which act correctly on the logical qubits will be very complex when encoded to apply to the physical spins.  A better approach is to manipulate the initial populations, so that the desired population pattern is shifted to the four states \ket{000}, \ket{001}, \ket{010} and \ket{011}, giving a much simpler relationship between logical and physical states \cite{Gershenfeld1997,Chuang1998b}.  The spin system is now in a pseudo-pure state, conditional on the first spin being in state \ket{0}, and logic gates can be implemented directly as long as they do not interchange the \ket{0} and \ket{1} states of this labelling spin \cite{Chuang1998b}.  This approach has been experimentally demonstrated to encode two logical qubits in a three-spin system \cite{Vandersypen1999, Dorai2000a}. Because it relies on naturally occurring patterns of identical populations, the approach is only applicable to homonuclear spin systems.

A more popular approach is \textit{temporal averaging by permutation} \cite{Knill1998}, which requires no additional qubits. In essence temporal averaging is similar to phase cycling, in that results from a number of similar experiments are averaged together, but here the experiments differ in the distribution of initial state populations. Since quantum logic gates and NMR readout are both linear processes, this is equivalent to performing a single experiment on an averaged input state. For example, on a two qubit system the experiment is run first on the thermal state and is then run preceded by each of the two cyclic permutations of the populations of the three excited state populations, leaving the ground state untouched in each case. This method works equally well with homonuclear and heteronuclear spin systems, as it makes no assumption about the pattern of populations beyond the lowest level having the highest initial population.

The most popular methods for preparing pseudo-pure states, however, are based on \textit{spatial averaging} \cite{Cory1996, Cory1997}, which is built around the use of magnetic field crusher gradients to dephase quantum states. The process in a two-spin homonuclear system can be easily understood using product operators \cite{Jones2011}. It normally begins by adjusting the relative populations of the two spins by partly exciting one of them and then applying a crusher gradient to remove off-diagonal terms.
\begin{equation}
\begin{split}
I_z+S_z
\xrightarrow{\makebox[2.5em]{$\scriptstyle60^{\circ}S_x$}}
&\textstyle I_z+\half S_z-\frac{\sqrt3}{2}S_y\\
\xrightarrow{\makebox[2.5em]{$\scriptstyle\textit{crush}$}}
&\textstyle I_z+\half S_z
\end{split}
\end{equation}
This is followed by a series of pulses, coupling periods, and a final crush gradient to convert $I_z$ to the right mixture of inphase and antiphase terms.
\begin{equation}
\begin{split}
I_z\xrightarrow{\makebox[2.5em]{$\scriptstyle45^{\circ}I_{x}$}}
&\textstyle\frac{1}{\sqrt2}I_z-\frac{1}{\sqrt2}I_y\\
\xrightarrow{\makebox[2.5em]{$\scriptstyle\textit{couple}$}}
&\textstyle\frac{1}{\sqrt2}I_z+\frac{1}{\sqrt2}2I_xS_z\\
\xrightarrow{\makebox[2.5em]{$\scriptstyle45^{\circ}I_{-y}$}}
&\textstyle\half I_z-\half I_x+\half 2I_xS_z+\half2I_zS_z\\
\xrightarrow{\makebox[2.5em]{$\scriptstyle\textit{crush}$}}
&\half I_z+\half 2I_zS_z
\end{split}
\end{equation}
where \textit{couple} indicates a delay of duration $1/2J$ for evolution under the pure spin--spin coupling Hamiltonian $\pi{J}\,2I_zS_z$. Note that the $\half S_z$ term is unaffected by the pulses and coupling terms, and comes through this stage unscathed. The process thus generates the correct final combination of terms for a pseudo-pure state. The coupling period can be implemented using spin echoes to refocus the Zeeman interactions, or alternatively such evolution can simply be tracked and the phases of subsequent pulses adjusted.  As the gradient pulses crush all off-diagonal terms, any rotations of the reference frame at the end of the process can simply be ignored, which significantly simplifies the implementation.

Whenever using sequential gradient crush sequences, it is necessary to guard against accidental gradient echoes, where two crush sequences cancel each other, causing crushed terms to be revived.  In homonuclear systems it is also important to avoid generating zero-quantum coherences, as these are not crushed by gradients. In heteronuclear systems zero-quantum coherence is not a problem and the simpler sequence \cite{Pravia1999}
\begin{equation}
\begin{split}
I_z+S_z &\xrightarrow{45^{\circ}(I_x+S_x)}
\xrightarrow{\textit{couple}}
\xrightarrow{30^{\circ}(I_{-y}+S_{-y})}\\
&\xrightarrow{\textit{crush}}
\textstyle\sqrt{\frac{3}{8}}\left(I_z+S_z+2I_zS_z\right)
\end{split}
\end{equation}
can be used. This sequence requires initially equal polarizations on the I and S spins, which can be achieved with a pulse applied to the higher polarization spin followed by a crush gradient, or with a more complex sequence \cite{Pravia1999} to average the two polarizations.

\subsection{Practical methods}
The methods of temporal averaging and spatial averaging can be extended from two spins to larger spin systems. Within temporal averaging, the naive \textit{exhaustive averaging} approach requires performing $2^q-1$ separate experiments on a system of $q$ qubits, and so is only practical for small systems. More efficient methods have been explored, combining non-cyclic permutations and unequal weights in the averaging process, permitting a pseudo-pure state to be prepared in a system of four homonuclear spins using a weighted sum of only five permutations rather than a naive average over 15 cyclic permutations \cite{Mori2005}. Alternatively, random permutations can be used to prepare approximate pseudo-pure states in very large systems \cite{Knill1998}.

In spatial averaging, the basic aim is to use unitary transformations to convert a thermal state to a state with an appropriate pattern of populations, and then apply crusher gradients to remove off-diagonal terms. This approach works well in fully heteronuclear systems, but difficulties arise in homonuclear systems, where zero-quantum coherence terms are unaffected by the crusher gradients. One solution to this is to use methods adapted from temporal averaging to perform qubit-selective crusher pulses, but as the number of experiments required doubles with every selective crush pulse applied \cite{Kawamura2010} this swiftly becomes impractical unless exhaustive averaging is replaced by a randomized process \cite{Bhole2020}.

Because of this it is not possible in homonuclear systems to simply apply a single unitary transformation (to assemble the correct population pattern) followed by a single crush pulse (to remove off-diagonal terms). Instead, it is necessary to alternate unitary and non-unitary transformations in a more complex pattern. The original method \cite{Cory1996, Cory1997} used the hand-designed sequence described above to generate the correct product operators with two crusher pulses \cite{Jones2011}. A more systematic approach uses \textit{controlled-transfer gates} \cite{Kawamura2010} to assemble the desired population pattern without ever generating zero-quantum coherences. This approach also has the advantage of extracting the largest possible amount of pseudo-pure state from a given initial state, and the method works equally well with heteronuclear spin systems or non-thermal initial states. However, the complexity of the sequences required means that they are rarely applied to systems with more than two spins.

Considerable simplifications to the networks required can be achieved if the single spin populations are first adjusted into a useful pattern, sacrificing optimal theoretical efficiency for practical simplicity. This can be achieved by applying selective excitations to a single spin and then applying a crush pulse to remove off-diagonal terms, as shown for a two-spin system in section~\ref{sec:simplePPS}.
\begin{figure}
\begin{center}
\includegraphics[width=74mm]{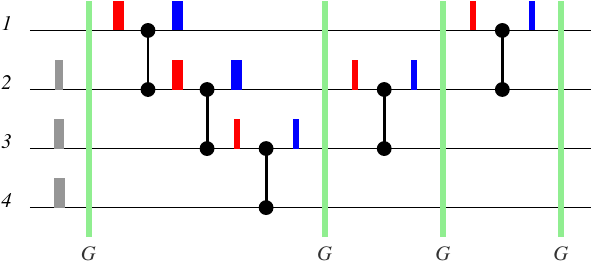}
\end{center}
\caption{A pulse sequence to generate a pseudo-pure state in a linear chain of four homonuclear spins such as crotonic acid. The spins are labelled 1 to 4 along the chain. The initial pulses applied to spins 2, 3 and 4, with $\theta_2=\arccos(1/2)=60^\circ$, $\theta_3=\arccos(1/4)\approx76^\circ$, $\theta_4=\arccos(1/8)\approx83^\circ$ respectively, followed by a crusher gradient ($G$), act to adjust the populations. Subsequent pulses are all $90^\circ$ (broad boxes) or $45^\circ$ (narrow boxes), with phases of $x$ for pulses before a coupling period, shown in red, and $-y$ for pulses after a coupling period, shown in blue. Coupling evolution for a time $1/2J$ under a single coupling, isolated using a spin echo, is shown as two circles connected by a line. The absolute phases of all pulses are unimportant, but the relative phases of red and blue pulses must be set correctly.} \label{fig:PPScrot}
\end{figure}
A particularly common approach with crotonic acid, a homonuclear four-spin system well approximated by a linear chain, is to adjust the populations along the chain to be in the ratios $8:4:2:1$, halving with every step down the chain, after which a simple sequence of just five controlled gates and three crush pulses can be used to generate a pseudo-pure state \cite{Kong2018}. A network for achieving this is shown in Figure~\ref{fig:PPScrot}; this network is very slightly simpler than the original, and uses the correct sign for the evolution under couplings. The size of the pseudo-pure state extracted can be enhanced by beginning the experiment with a non-thermal state in which populations are enhanced by nuclear Overhauser effects \cite{Bhole2020}.

Many other methods for generating pseudo-pure states have been explored, in particular combining  temporal averaging and spatial averaging methods to get the advantages of both \cite{Mahesh2001a, Knill1998, Kawamura2004, Zheng2019}, and using highly entangled states \cite{Knill2000} or singlet states \cite{Roy2010a}. It is in general much simpler to produce states which are almost pseudo-pure than fully pseudo-pure states, and several techniques for doing this have been described \cite{Fung2001a, Peng2004, Fung2004}.

\subsection{Fidelities}
Given the emphasis on optimal control through computer search in earlier sections, it might seem odd that the preparation of pseudo-pure states remains dominated by hand-designed approaches. One reason for this is that the conventional fidelity formulae are not easily applicable in this case, as they tend to assume that either one of the states involved is pure, or that the quantum evolution is unitary, or both. Since a pseudo-pure state is a highly mixed state, and must be prepared by a non-unitary process, great care must be taken.

Suppose it is desired to prepare a pseudo-pure state corresponding to the pure state \ket{00} in a two-spin system. It might seem that
\begin{equation}
\bra{00}\rho\ket{00}
\end{equation}
would provide a suitable fidelity expression for a general state $\rho$. However, this expression simply identifies the size of the component of $\rho$ which is parallel to \ket{00}, and is entirely insensitive to any other property. Thus, for example, the two states
\begin{equation}
\rho_1=\begin{pmatrix}\half&0&0&0\\0&\textstyle\frac{1}{6}&0&0\\0&0&\textstyle\frac{1}{6}&0\\0&0&0&\textstyle\frac{1}{6}\end{pmatrix}
\end{equation}
and
\begin{equation}
\rho_2=\begin{pmatrix}\half&0&0&0\\0&\half&0&0\\0&0&0&0\\0&0&0&0\end{pmatrix}
\end{equation}
would give precisely the same result, even though $\rho_1$ clearly is a pseudo-pure state, and $\rho_2$ clearly is not.

Similar issues arise if the naive mixed state fidelity, $\text{tr}(\rho\sigma)$, is used to compare a general state $\rho$ with a target pseudo-pure state
\begin{equation}
\sigma=(1-p)\mathbf{1}/4+p\ket{00}\bra{00}.
\end{equation}
As the trace operation is linear this is a weighted sum of contributions from the pure component, which leads to the problems discussed above, and from the maximally mixed component, which reduces to $\text{tr}(\rho)/4$, which is simply equal to $1/4$ for any properly normalised density matrix.

For these reasons conventional fidelity functions are rarely useful when designing networks to prepare pseudo-pure states. It is possible to fall back to the Uhlmann--Jozsa fidelity, or to other measures of infidelity, such as $||\rho-\sigma||$ for some suitable matrix norm, but while these are suitable for testing whether two states are identical they might not be particularly useful for comparing the quality of two imperfect matches to the desired state. For example, a pseudo-pure state with sub-optimal effective purity is likely to be more useful for practical purposes than a state of the wrong form, even if this is formally closer to the desired state.

\section{Closed-loop control}\label{sec:CLC}
All the methods described so far have been examples of \textit{open-loop control}, in which the control sequence is designed on a computer using a description of the physical system, and then simply implemented on it. The underlying physical system is not used in the design of the control sequence, except possibly in some final calibration experiments. A radically different approach is provided by \textit{closed-loop control}, in which the physical system itself is used as the principal design tool. Rather than calculating fidelities, which is computationally expensive, the actual state-to-state fidelity is measured experimentally, and the control parameters are adjusted to optimize it.

Since being proposed as a route for controlling quantum systems with laser pulses \cite{Judson1992}, the method has been widely explored  \cite{Bardeen1997, Assion1998, Levis2001, Rabitz2002, Zhu2003a}. The approach has two major advantages over open-loop control, both of which arise from the use of the quantum system to study the effects of the control sequence. Firstly, simulating the control sequence using an explicitly quantum mechanical physical system avoids the exponential complexity blow-up inherent in classical simulations \cite{Feynman1982}, by in effect using the quantum system to simulate its own behaviour \cite{Toffoli1982}. Secondly, using the system itself allows the true parameters actually describing the system to be used, rather than approximate measured values, and uses the control fields actually applied, rather than those requested. If the initial state can be easily prepared and the final state easily characterised, then measuring state-to-state fidelities is straightforward, and with the technologies normally used it is possible to apply thousands or even millions of trial control sequences to the system every second. More recently it has been suggested that closed-loop feedback can be combined with open-loop GRAPE control to get the best of both approaches \cite{Porotti2023}.

Closed-loop quantum control has been less frequently applied in NMR, although it has been used in the design of an NMR gyroscope using optical readout \cite{Zhang2020} and within NMR QIP for the preparation of Bell states \cite{Yang2020b} and for quantum metrology \cite{Yang2020a}. The achievable repetition rate is usually quite slow with NMR, as the long relaxation times limit the rate at which initial states can be prepared. The ensemble nature of the NMR readout process is an advantage, but this is also the case in some other implementations.

The most important weakness of closed-loop optimization is that it is really only suitable for state-to-state fidelities, and cannot easily be generalized to design true unitary transformations. To do the latter requires finding the state-to-state fidelity for an exponentially large number of initial states that span the basis of dimension $2^q$ for a system with $q$ qubits. This is not quite as bad as performing full quantum process tomography \cite{Chuang1997, DAriano2001}, but it remains a very challenging process for systems with more than a few qubits.

\subsection{Randomized benchmarking}
Although quantum process tomography takes too long to be useful in closed-loop control, it has been demonstrated for assessing the performance of control sequences in simple cases \cite{Childs2001, Weinstein2004}. As for quantum state tomography, methods have been developed to make the process more efficient \cite{Wu2013a, Shukla2014a, Maciel2015, Wang2016, Gaikwad2018}, but it remains a challenging task, and it is desirable to find some simpler quality measure.

One popular approach is randomized benchmarking \cite{Knill2008, Hines2023}, which aims to estimate the relevant fidelity of a set of quantum logic gates for implementing complex quantum networks by applying long sequences in random orders. Note that the method cannot be applied to characterize a single gate, and more general questions have been raised about the meaning and value of such measurements \cite{Proctor2017}, especially in the presence of correlated (non-Markovian) errors \cite{Edmunds2020, FigueroaRomero2021}.

The method has been demonstrated on NMR implementations of three qubit \cite{Ryan2009} and four qubit \cite{Xin2018a} systems, and has also been used to monitor calibration errors in electron spin resonance \cite{Park2016}. It is possible to combine randomized benchmarking with partial quantum process tomography when more detailed information is desired \cite{Jiang2018a}. Other methods for estimating average fidelities have also been explored \cite{Lu2015}.

\section{Refocusing networks}\label{sec:WHSE}
The use of optimal control methods opens up very considerable freedom in the design of experiments to implement quantum algorithms. Conventionally an algorithm will be written as a network of logic gates, which can then be compiled into a longer network of simpler one- and two-qubit logic gates, forming a universal set \cite{Barenco1995}. All that is then necessary is to implement a small number of logic gates, spanning the universal set.

This might not, however, be the best way to proceed, and it is common for experimentalists to design optimal control sequences which directly implement more complex gates, such as the Toffoli gate \cite{Atia2014}, or more exotic gates such as the partial SWAP \cite{Violaris2021}. Similarly, one can design a control sequence which implements a small network of more basic gates in one go, or even implement an entire algorithm in one control sequence \cite{Li2017d}. This final approach can, however, become illegitimate, with all the work of the algorithm actually being done by the compiler \cite{Smolin2013}.

At the other extreme it can be useful to restrict oneself to using only single-qubit gates and free evolution under the drift Hamiltonian \cite{Peterson2020}, essentially equivalent to using pulses and delays in conventional NMR. This greatly simplifies the GRAPE problem, as it is only necessary to design gates which act selectively on individual spins, or on groups of spins, leading to much shorter pulses than those designed to implement controlled logic. Two-qubit gates are implemented through periods of free evolution under a Hamiltonian containing only single spin $z$ terms and two-spin $zz$ interactions. During this time no RF is applied, reducing the scope for error. As well as being demonstrated in NMR systems containing 4, 7, and 12 qubits, simulations have been performed in fictional square two-dimensional lattices containing 16, 36, and 100 qubits, suggesting that the method can be scaled to very large systems \cite{Peterson2020}.

Within this approach it becomes very important to find methods for designing efficient spin echo sequences that sculpt the drift Hamiltonian into a more desirable form. The conventional NMR approach of nested spin echoes is adequate for small systems, but becomes unwieldy above a handful of spins \cite{Jones2000a}. Fortunately far more efficient methods exist. These methods are all designed to select or to rescale couplings within an extended network as efficiently as possible, while simply refocusing all single spin offset frequencies (chemical shifts). When single qubit $z$ rotations are required, for example to turn coupling gates into controlled-phase gates \cite{Jones1998b}, this can be easily achieved: applying two $180^\circ$ rotations around axes in the $xy$-plane that are separated by a phase angle $\delta$ is equivalent to performing a $z$-rotation through an angle $2\delta$,
\begin{equation}
180^\circ_{\phi_2}180^\circ_{\phi_1}=2(\phi_2-\phi_1)_z,\label{eq:phiz}
\end{equation}
which can be interpreted as an Aharonov--Anandan phase \cite{Suter1988}. This approach is far more convenient than the conventional composite $z$ rotation \cite{Freeman1981}, as it can be combined with the refocusing network by simply changing the relative phase of two refocusing pulses.

The most basic task in Hamiltonian sculpting is to refocus all the chemical shifts and all but one of the couplings, so that the effective evolution is under the single retained coupling. In a two-spin $IS$ system this can be achieved by applying $180^\circ$ pulses to both spin $I$ and spin $S$ at time $t/2$, half way through the evolution period $t$. For completeness, a second pair of $180^\circ$ pulses should be applied at the end of the evolution period, although in conventional NMR this is frequently omitted.

The way to understand this spin echo \cite{Hahn1950} is that $180^\circ$ pulses reverse the sign of the chemical shift evolution, so that evolution in opposite directions for two equal times causes it to cancel overall, but as the pulses are applied to both spins the $zz$ coupling is reversed twice, and so left unchanged. In a three spin $ISR$ system it becomes necessary to add $180^\circ$ pulses at times $t/4$ and $3t/4$, dividing the individual evolution times in two again. In a four spin $ISRT$ system these times would be subdivided yet again, with four $180^\circ$ pulses applied to spin $T$ at times corresponding to odd multiples of $t/8$. Clearly the process can be extended to any number of spins, but this approach will result in an exponential growth in both the number of time periods and the number of $180^\circ$ pulses as the number of spins is increased.

\subsection{Walsh--Hadamard patterns}
Fortunately this naive approach is not the best way to tackle large numbers of spins. Instead, more efficient refocusing schemes can be devised \cite{Jones1999, Leung2000, Leung2002}, based on the properties of Hadamard matrices, and requiring a number of time periods that scales only linearly with the number of spins, and a number of pulses that scales only quadratically. These methods are best described using Walsh--Hadamard matrices, where each row is a Walsh function \cite{Beauchamp1984}. These are only defined for dimensions equal to a power of two, while more general Hadamard matrices can be defined for most multiples of 4 \cite{Jones1999}. They differ from the standard Hadamard matrices used in QIP \cite{NCbook} in the rows not being normalised, and the ordering of the rows being different.

A Walsh function $W^N_n$ is defined by a vector with length $N$ equal to a power of 2 and with all the entries set to $\pm1$.  For every strictly positive integer $n<N$ the vector $W_n^N$ has half the entries set to $+1$ and half set to $-1$, with the entries arranged to create $n$ regularly spaced sign changes along the row, while for the special case of $W_0^N$ all the entries are $+1$, so there are no sign changes, as expected for $n=0$. For the case $N=4$ the Walsh--Hadamard matrix contains the four rows listed in Figure~\ref{fig:WHpulses}.
\begin{figure}
\begin{center}
\includegraphics{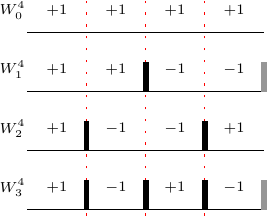}
\end{center}
\caption{The four Walsh functions $W^4_j$ and the patterns of $180^\circ$ pulses which generate them. Note that pulses are applied whenever the function changes sign. The pulses shown in grey are not necessary to generate the desired modulation, but are required to return the effective Hamiltonian to its initial sign.} \label{fig:WHpulses}
\end{figure}

From now on the superscript value of $N$, which specifies the number of columns, will be dropped, leaving only the subscript $n$ indicating the number of sign changes. The value of $N$ is specified implicitly, being equal to the smallest power of 2 larger than the highest Walsh number considered. The Walsh functions can be considered as digital equivalents of sine and cosine functions, and are sometimes called sal (for functions with odd parity around the middle) and cal (for functions with even parity) \cite{Beauchamp1984}, but treating them as a single basis set is more useful here.

A single spin $z$ interaction can be refocused by ensuring that its pattern of evolutions corresponds to a Walsh function other than $W_0$, which can itself be achieved by applying a $180^\circ$ pulses at points corresponding to sign changes, as shown in Figure~\ref{fig:WHpulses}. The $zz$ coupling between two spins will evolve with a pattern described by the product of the two corresponding Walsh functions, which is itself a Walsh function given by
\begin{equation}
W_m\circ W_n=W_{m\oplus n}
\end{equation}
where the $\circ$ symbol indicates element-wise multiplication, sometimes called the Schur product \cite{Lynn1964}, and the $\oplus$ symbol indicates bitwise addition modulo 2. For example, $W_2\circ W_3=W_1$, which is easily verified directly.

Since $m\oplus n$ equals 0 if and only if $m=n$, this means that all couplings will also be refocused unless two spins experience the same pulses, in which case the coupling will be retained at full strength. Thus the optimal way to sculpt the drift Hamiltonian to isolate a single coupling is to assign the two coupled spins to the pattern $W_1$ and all other spins to successively higher numbered $W_n$.

\subsection{Time optimal refocusing}
The procedure above can be used to assemble a set of one-qubit $z$ and two-qubit $zz$ interactions by isolating each coupling in turn and implementing single qubit rotations using equation~\ref{eq:phiz} to choose appropriate relative phases for two $180^\circ$ pulses. However, although each individual step is optimal this will not normally achieve the desired evolution in the shortest possible time, as it is sometimes possible to retain several different coupling interactions in parallel.

The simplest case where this cannot be achieved is provided by a system of three coupled spins. Here it is simple to design refocusing sequences which retain any one of the three couplings between the spins, while refocusing the other two, but it is impossible to retain two couplings while refocusing the third. Thus to achieve coupling evolution under two couplings it is necessary to perform separate evolutions under each coupling, applying two refocusing sequences back-to-back. In larger spin systems, however, it is possible to select certain subsets of couplings: for example, in a system of four coupled spins it is easy to simultaneously retain couplings between spins 1 and 2, and between spins 3 and 4, while refocusing everything else.

Finding the time-optimal refocusing pattern is not a trivial problem, but it can be accomplished using methods from linear programming \cite{Bhole2020a}. The method starts by assigning spins to Walsh patterns with numbers given by successive powers of 2. This guarantees that every one- and two-spin interaction will be assigned to a \textit{unique} Walsh patten, and so they can all be controlled independently. Linear programming then seeks a set of evolution times which achieves the desired overall evolution in the shortest possible time, subject to the constraint that all individual times must be non-negative. In practice it is more stable to use time symmetrised solutions, which automatically remove all single qubit terms, and then reintroduce these through phase shifts \cite{Bhole2020a}.

Linear programming is a practical solution for systems up to around 20 spins, after which the time required to find solutions, which grows exponentially with the number of spins, becomes impractical. This is unlikely to prove an important restriction as NMR QIP systems larger than this appear impractical for other reasons \cite{Jones2000a}. However if necessary it is possible to use approximate methods to locate near-optimal solutions in a much shorter time, with only polynomial time scaling, and this has been demonstrated for simulated systems of up to 125 spins \cite{Bhole2020a}.

\subsection{Engineered networks}
If very large devices are ever implemented using NMR QIP or related techniques then it is likely that these will be engineered systems, rather than natural molecules. A simple model is to assume that the spins form a two-dimensional square array, with couplings only between near neighbours \cite{Peterson2020}. For the case of a square array with only nearest-neighbour couplings there exists a simple constructive algorithm for designing near-optimal refocusing networks in a time which is \textit{linear} in the number of spins, and so scalable up to arbitrary sizes \cite{Tsunoda2020}. The resulting patterns are never worse than a factor of two slower than the true time-optimal solutions, and are robust to the presence of next-nearest-neighbour couplings. Related ideas have been explored in superconducting qubits \cite{Le2023}.

\section{Dynamical decoupling}\label{sec:dd}
Dynamical decoupling \cite{Viola1998, Viola1999, Morong2023} refers to a family of methods for removing unwanted interactions between a quantum system and its environment, ultimately built upon the Hahn spin echo \cite{Hahn1950} and methods for coherent averaging \cite{Haeberlen1968}. Although the term sounds very similar to decoupling in NMR, it differs from it in one central way: the control pulses are applied to the \textit{system}, rather than to the \textit{environment}. The ultimate aim is to retain the state of a qubit unchanged as far as possible, producing a reliable memory \cite{Souza2011b}.

Both decoupling and dynamical decoupling seek to remove unwanted interactions by applying control sequences which cause spin echoes. If the interactions were static and local then a single spin echo would suffice, but noise can cause these interactions to fluctuate, while additional strong interactions within the environment can cause information to spread out beyond the original spin pair. For this reason it is necessary to apply spin echoes repeatedly, ideally rapidly compared with the fluctuation rate and compared with the sizes of the interactions within the environment. In conventional NMR the environment is frequently dominated by spins of a different nuclear species to the system, and it is practical to apply the control sequences to the environment spins, decoupling them from the system \cite{Levitt1981, Waugh1982, Waugh1982a, Shaka1987a}.  In general, however, the environment can be far more varied, and may be uncontrollable, in which case control pulses have to be applied to the system itself. This is familiar within conventional NMR as the CPMG spin echo train \cite{Carr1954, Meiboom1958}.

This conceptual difference leads to significant practical differences. Because pulses are applied to the system itself there is a danger of dephasing due to inhomogeneity in the RF field. For this reason it is important that the $180^\circ$ pulses are designed to be as accurate as possible in the presence of systematic errors, and that they are designed to perform well as general rotors, and not just as inversion pulses as is the case for conventional decoupling. Similarly, any phase sequence which is applied must ensure that the quantum state is returned as accurately as possible to its original state at the end of the sequence. Note that even in the absence of errors the qubit will only return to its initial state at certain points in the decoupling cycle, and so the quality of dynamical decoupling can only be properly assessed at the end of a cycle, or at least of a shorter sub-cycle, just as the quality of CPMG refocusing should only be considered after an even number of $180^\circ$ pulses.

There are three significant features that need to be considered when designing a dynamical decoupling sequence:  the spacing between the $180^\circ$ pulses, the design of individual pulses, and any phase modulation which is applied to successive pulses. The choice of spacing depends on the noise spectrum of the interaction to be refocused. If the interaction is static then it suffices to apply a single pair of spin echoes, each of which refocuses the undesired interaction during its own echo period and which in combination act as an identity operation. If, however, the interaction is time varying, for example due to diffusion  \cite{Stejskal1965} or chemical exchange \cite{Luz1963}, then the interaction is only effectively suppressed if the $180^\circ$ pulses are applied \textit{rapidly} in comparison with the variation \cite{Baldwin2009}.

This dependence of suppression of an interaction on pulse spacing can be used to measure the spectrum of the interaction, or to distinguish between systems according to their sizes \cite{Morris1992, Jones1997b}, but it may simply be desired to suppress the interaction as far as possible. The obvious approach is to apply the echoes as fast as possible, culminating in continuous dynamical decoupling, in which pulses are applied back to back, just as they normally are in conventional decoupling. In practice the performance of rapid dynamical decoupling initially improves as the pulse spacing is reduced, but beyond a certain point the damaging effects of errors in the pulses dominate over improved suppression, and the best performance is normally seen for some small but non-zero pulse spacing, as discussed in section~\ref{sec:RDD}. It can also be desirable to keep space between pulses in order to reduce the total RF power necessary \cite{Souza2012}.

Surprisingly, the best performance is not always seen with evenly spaced echoes. Uhrig dynamical decoupling, discussed in section~\ref{sec:UDD}, involves a carefully chosen set of unequal pulse spacings. This result was described as ``the first case of this framework [QIP] enabling magnetic resonance (MR) applications'' \cite{Jenista2009}, and is certainly one of the most relevant insights from QIP for conventional NMR.

\subsection{Rapid dynamical decoupling}\label{sec:RDD}
\begin{figure*}
\begin{center}
\includegraphics[width=16cm]{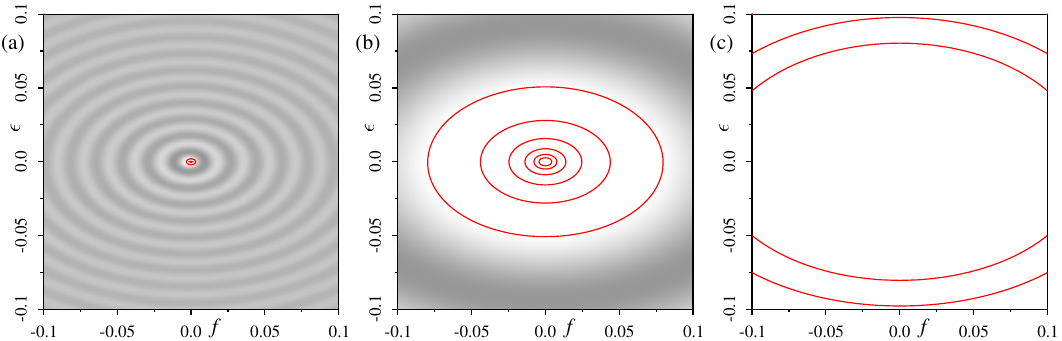}
\end{center}
\caption{Simulated performance of three different approaches to rapid dynamical decoupling: (a) CPMG, (b) XY-4, and (c) KDD4. The plots show a fidelity measure averaged over initial states along $x$, $y$, and $z$, appropriate for a qubit memory, after a total of 180 spin echoes, in the presence of both pulse strength errors $\epsilon$ and off-resonance effects $f$. Fidelity contours are drawn at six infidelity levels, logarithmically spaced between $10^{-1}$ and $10^{-6}$, and control errors cover a range of $\pm10\%$, parameterised as fractions of the driving field strength \cite{Jones2013}. CPMG only performs well when control errors are negligible, reflecting the poor preservation of magnetisation perpendicular to the control fields, but the XY-4 sequence is a vast improvement. A similar gain is seen for KDD4 (that is, using the Knill pulse phases as an inner phase modulation cycle with XY-4 outside this to give a twenty-step cycle) where only the two highest contours are visible. If the Knill pulse is replaced by a nine pulse sequence, as described in the text, the fidelity is above the highest contour (infidelity below $10^{-6}$) across very nearly the entire range considered (not shown). }\label{fig:RDD}
\end{figure*}
With rapid dynamical decoupling it is important to minimise the effects of systematic errors in the driving fields on the state of the system, through a mixture of phase sequences and composite pulses \cite{Souza2011b}. For simplicity I will consider the case of a single spin in the presence of phase noise, due to variations in the local magnetic field strength. If these variations arise from $B_0$ inhomogeneity they will be static, unless molecular motion causes them to fluctuate. If they arise from couplings to other spins, then fluctuations can also occur due to relaxation of the coupling partners. Whatever the cause, the effect can be modelled as an additional $z$ interaction, which varies both across the ensemble and in time. Spin states along $z$ will be unaffected, but states in the $xy$ plane will be dephased by the interaction. Variation across the ensemble can be suppressed by a simple spin echo, but variation in time requires a series of echoes, naively with a spacing short compared with the timescale over which the interaction varies.

This long sequence of spin echoes requires a correspondingly large number of $180^\circ$ pulses, and if these are not perfect then errors, such as pulse strength or duration errors and off resonance effects, will build up. However, it is a remarkable feature of the CPMG sequence that these errors largely cancel out on even-numbered echoes for initial states aligned along the pulse direction. Specifically, if the $180^\circ$ pulses are applied along $x$, corresponding to \NOT\ gates, and the initial spin state is also aligned along $x$, then a single spin echo gives a signal which is not at full strength but instead is reduced quadratically by both pulse strength errors and off resonance effects. If the initial spin state is aligned along $y$ or $z$ then the spin echo causes the state to be inverted, once again with quadratic errors. On the second echo, however, the error for a state initially along $x$ is reduced to fourth order, while states along $y$ or $z$ are returned to their original direction but retain the quadratic error terms. Related effects are seen in spin locking experiments \cite{Freeman1971}.

For this reason a CPMG sequence is much better at preserving qubits in one direction (aligned with the control field) than any other. If the pulses are instead applied alternately along $\pm x$ then initial states along $y$ now exhibit only fourth order error dependence, while $x$ and $z$ show quadratic errors. Note that states initially along $z$ are naturally invulnerable to phase noise, and so the effects of the CPMG sequence are purely damaging in this case. More complex behaviour can arise in more realistic situations \cite{Franzoni2005, Li2007}, but the broad conclusions are unaffected.

One solution is to use a more complex phase sequence, such as XY-4 \cite{Maudsley1986, Gullion1990}, in which the $180^\circ$ pulses are applied alternately along $x$ and $y$. In this case the initial state is only restored after every fourth pulse, but the error tolerance is greatly improved, with fourth order errors for initial states along $x$ and $y$ and sixth order errors for initial states along $z$. As a consequence, XY-4 dynamical decoupling is moderately effective at preserving all initial states even for large numbers of echoes  \cite{Souza2011b}, as shown in Figure~\ref{fig:RDD}.

To gain further improvements one could use a longer sequence, such as XY-8 \cite{Ryan2010, Ahmed2013, Zhang2014}, but an alternative approach is to replace the $180^\circ$ pulses with composite pulses \cite{Borneman2010}. For use with conventional decoupling, composite pulses should be optimised to act as inversion pulses, but for dynamical decoupling it is important that the pulses act as universal rotors, sometimes called class-A composite pulses, which perform well for all initial states \cite{Levitt1986}. A particularly useful group of composite $180^\circ$ pulses is obtained by using an odd number of $180^\circ$ pulses with carefully chosen phases, particularly when these phases are chosen to be time symmetric \cite{Jones2013}. Among such pulses the sequence
\begin{equation}
180_{30}\,180_0\,180_{90}\,180_0\,180_{30}
\end{equation}
which is sometimes called the Knill pulse \cite{Souza2011b, Ryan2010} is particularly suitable; note that as usually described this does not implement a \NOT\ gate, but this can be remedied by offsetting all the phases by $210^\circ$ \cite{Jones2013} to give
\begin{equation}
180_{240}\,180_{210}\,180_{300}\,180_{210}\,180_{240}.
\end{equation}
This pulse performs a \NOT\ gate with tolerance of both pulse strength errors and off-resonance effects, and unlike some alternatives has good tolerance of simultaneous errors \cite{Jones2013}.

Composite pulses of this kind can be used in two different ways. The obvious approach is to replace each $180^\circ$ pulse in a decoupling sequence with a composite pulse, but for the Knill pulse this increases the number of pulses used, and thus the total power applied, by a factor of five, unless the spacing between the refocusing pulses is increased to compensate. Alternatively, the spacing can be left unchanged, and the phases of the Knill pulse imposed as a phase cycle. This must then be combined with XY-4 phase cycling to get a complete cycle of length 20. This second approach, sometimes called Knill dynamical decoupling \cite{Souza2011b, Wang2021, Drmota2023}, is the most effective.

This final approach could in principle be extended by using even more effective composite pulses, such as the sequence of nine $180^\circ$ pulses with phases
\begin{equation}
\alpha,\,\beta,\,\beta,\,\beta-\pi,\,2\beta-2\alpha,\,\beta-\pi,\,\beta,\,\beta,\,\alpha
\end{equation}
where
\begin{equation}
\beta=2\alpha+\arccos[-(1+2\cos\alpha)/2]
\end{equation}
and
\begin{equation}
\alpha=-\arccos[(4-\sqrt{10})/4],
\end{equation}
so $\alpha\approx-77.9^\circ$ and $\beta\approx-20.6^\circ$, which has exceptional tolerance of both pulse strength errors and off-resonance effects \cite{Jones2013}. However its performance in practice has yet to be explored.

It is also possible to combine dynamical decoupling with optimal control \cite{Borneman2010, Tabuchi2017}, replacing hard pulses with shaped pulses; preliminary explorations suggest that this will be a promising approach \cite{Yang2022a}.

\subsection{Uhrig dynamical decoupling}\label{sec:UDD}
The calculations shown in Figure~\ref{fig:RDD} assumed that the dephasing being refocused is unknown but constant during the decoupling period, or equivalently that it varies across the ensemble of spins being observed but does not vary in time. If this were in fact the case it would not be necessary to use rapid decoupling, as a single spin echo would be sufficient to refocus such static dephasing. It might be desirable to use two spin echoes, in order to restore the original state, or to use four spin echoes to permit the use of the XY-4 phase sequence, but there is no reason to perform large numbers of echoes.
\begin{figure*}
\begin{center}
\includegraphics[width=16cm]{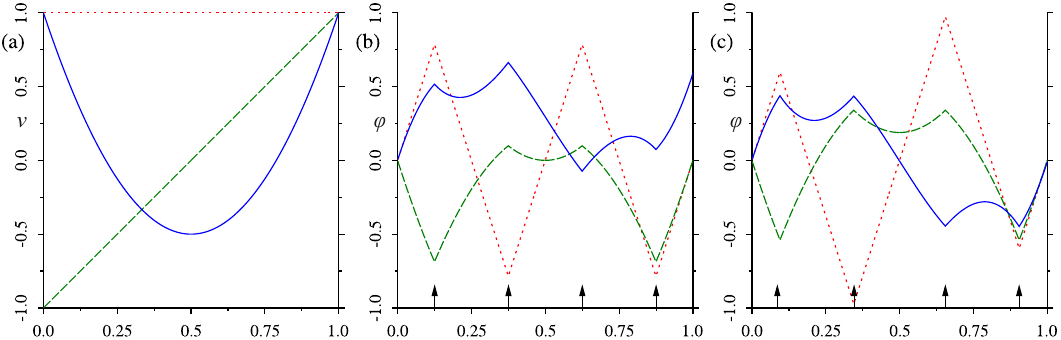}
\end{center}
\caption{Simulated action of spin echoes with conventional and Uhrig spacings. Plot (a) shows three different models for the offset frequency which needs to be refocused: the red dotted line shows a constant offset, the green dashed line shows an offset which varies linearly with time, and the blue solid line shows quadratic variation. Plot (b) shows the accumulated phase for a conventional echo sequence, with $\pi$ pulses at the positions indicated by black arrows causing the direction of phase accumulation to be reversed. The constant and linear offsets are refocused but an overall phase remains from the quadratic offset. Plot (c) shows the accumulated phases with Uhrig pulse spacing, and all three offsets are now refocused.}\label{fig:UDDechoes}
\end{figure*}

This changes if the dephasing varies with time. The original CP (method B) \cite{Carr1954} and CPMG \cite{Meiboom1958} sequences were designed to tackle losses due to diffusion within magnetic field gradients, and in this case the conventional approach is to apply evenly spaced echoes ar rapidly as possible. However, dephasing noise can arise for a variety of reasons, and the assumption that even spacing is always best is incorrect. An early result showed that concatenated dynamical decoupling could be more effective than the standard periodic approach \cite{Khodjasteh2005, Khodjasteh2007}, but other than placing some pulses back-to-back this is still built around even spacings, and ultimately achieves better performance by applying pulses very rapidly.

\begin{figure}
\begin{center}
\includegraphics{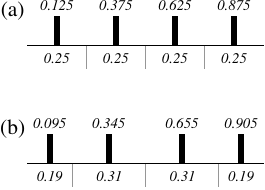}
\end{center}
\caption{How to divide a time interval by 4 pulses for (a) periodic dynamical decoupling, and (b) Uhrig dynamical decoupling. Pulses are shown with their time positions indicated above; the corresponding individual echoes and their lengths are shown below.} \label{fig:UDDtimes}
\end{figure}

A more radical departure is Uhrig dynamical decoupling \cite{Uhrig2007, Uhrig2008}, which starts by assuming that the number of refocusing pulses will be small, and asking how best to separate them. The original result assumed a particular model for dephasing noise, but was subsequently shown to apply more generally for slowly varying noise \cite{Lee2008, Uhrig2009}. If a total time period $T$ is to be divided into spin echoes by $n$ pulses then the optimal times for these pulses are given by
\begin{equation}
t_j=T\sin^2\left(\frac{\pi j}{2n+2}\right)\label{eq:UDDtimes}
\end{equation}
where $j$ runs from 1 to $n$. For the case $n=2$ this places pulses at $T/4$ and $3T/4$, reproducing the standard periodic pattern, but for higher $n$ the pulses are concentrated towards the start and end of the time period. The case $n=4$ is shown in Figures \ref{fig:UDDechoes} and \ref{fig:UDDtimes}: for a periodic pattern such as CPMG or XY-4 pulses are placed at odd multiples of $T/8$, producing 4 echoes of length $T/4$, but for Uhrig decoupling the first and last echoes are shortened to $0.19T$ while the middle echoes are lengthened to $0.31T$.

The conventional approach is designed to refocus a constant offset, but will also refocus a frequency offset which varies linearly with time: indeed a set of symmetrically arranged pulses will refocus any offset variation which is an odd function of time. However an offset which varies quadratically with time is not refocused, but results in an overall buildup of phase. (The offset functions shown in Figure~\ref{fig:UDDechoes} are shifted Legendre polynomials, which are mutually orthogonal, and so the quadratic function is a purely quadratic variation, with no constant or linear term.) By contrast, choosing the pulse spacing according to Uhrig's formula leads to all three terms being refocused.

Uhrig decoupling can be understood by considering the noisy dephasing Hamiltonian in a toggling frame generated by the pattern of $180^\circ$ pulses. The noise can be decomposed into components of different frequencies, and while the static component will be cancelled by any pattern of echoes, other frequencies will only be directly cancelled by echoes which are stroboscopic with that frequency. Uhrig decoupling considers the overall degree of suppression for the whole sequence of echoes as a function of frequency, and expands the response as a Taylor series around zero-frequency. It can be shown \cite{Jenista2009, Szwer2010} that the times in equation~\ref{eq:UDDtimes} set all the leading terms in this expansion to zero, resulting in good suppression in a broad band around zero-frequency. The truly optimal approach depends on the precise spectrum of the relevant noise sources \cite{Chen2022}. The original analysis assumed instantaneous refocusing pulses, but the effects of finite pulse width can be included \cite{Pasini2011}

Uhrig decoupling has been demonstrated experimentally in a range of systems, including NMR \cite{Jenista2009, Roy2011, Chakraborty2015, Schirmer2017, Singh2017, Bhattacharyya2020}, electron spin resonance \cite{Du2009, Rong2011}, and trapped ions \cite{Biercuk2009, Biercuk2009a}, and in general the expected benefits are seen. One significant disadvantage is that all pulses are applied with the same phase, and thus the method suffers from the same sensitivity to pulse errors as seen in CPMG, although the number of pulses used can be significantly smaller. This is not always important in conventional NMR, as the initial state of the magnetization is often known beforehand, and the pulses can be aligned with that state, but it is a more significant issue for quantum memories, which must work for all states. It is, of course, possible to use composite pulses to tackle this, but this must be done by simply replacing each pulse in the Uhrig sequence by a composite pulse, rather than using the pulse design to create an inner phase cycle. Optimized pulses have also been specifically designed for use with Uhrig dynamical decoupling \cite{Pasini2008a, Pasini2008, Karbach2008, Fauseweh2012}. These ideas are now being combined with methods from shaped pulse design to develop excitation sequences which are robust to time-varying interactions \cite{Li2021a}.

\section{Conclusions}
As was predicted in the early days of the field \cite{Warren1997, Gershenfeld1997a, Jones2000a, Jones1999b, Cummins2000a}, NMR has not led to a general scalable implementation of a quantum computer, and in recent years it has ceased to lead the field in the implementation of small demonstration devices. Superconducting quantum computers \cite{Ripoll2022} are now available with many more qubits than are available in NMR implementations \cite{Arute2019etal, Castelvecchi2023}, while ion trap implementations can beat NMR is speed and precision \cite{Bermudez2017, Schaefer2018}, and reconfigurable atom arrays have been used to demonstrate multiple logical qubits using advanced error correcting codes \cite{Bluvstein2023}. Despite this NMR implementations can still in practice compete with other approaches in at least some cases \cite{Greganti2021}.

As was also predicted the main role of NMR QIP has become a route for technology transfer, in both directions \cite{Jones1999b, Jones2001}. The long-standing emphasis within conventional NMR on composite pulses and shaped pulses has led to these ideas being transferred into other fields where precise quantum control is important \cite{Schaefer2018}. Particular methods have been developed within NMR QIP, among which the GRAPE algorithm stands out as the most generally useful approach. GRAPE has also been used to design shaped pulses for applications in conventional NMR, and it is gradually becoming understood within the NMR community that shaped pulses designed with optimal control can out-perform those designed by more conventional heuristic processes. The application of these ideas to electron spin resonance has been slower, reflecting the much greater complexity of implementing arbitrary waveforms at these high frequencies and short pulse widths \cite{Endeward2023}, but initial experience has proved promising \cite{Spindler2012}.  A second important area is of developments in decoupling arising from the field of dynamical decoupling, and especially Uhrig dynamical decoupling, in the presence of time-varying interactions.  Although the field is unquestionably becoming quieter, interesting and important things still remain to be done.

\section*{Acknowledgments}
I thank Andy Baldwin and Gaurav Bhole for many helpful conversations. This review is ultimately based on a talk delivered to the Oxford University Quantum Information Society, and was influenced by questions and comments from several NMR experts who attended.


\bibliography{ProgressV3}

\section*{Glossary}
\setlist[description]{font=\normalfont\itshape}
\begin{description}
\item[BB1:] broad band number one
\item[BFGS:] Broyden--Fletcher--Goldfarb--Shanno
\item[BURP:] band-selective uniform response pure-phase
\item[CCCP:] concatenated composite pulse
\item[CP:] Carr--Purcell
\item[CPMG:] Carr--Purcell--Meiboom--Gill
\item[CRAB:] chopped random basis
\item[ENDOR:] electron nuclear double resonance
\item[GOAT:] gradient optimization of analytic controls
\item[GRAPE] gradient ascent pulse engineering
\item[GRAWME:] gradient ascent without matrix exponentiation
\item[HMQC:] heteronuclear multiple quantum coherence
\item[HSQC:] heteronuclear single quantum coherence
\item[L-BFGS:] limited memory Broyden--Fletcher--Goldfarb--Shanno
\item[MR:] magnetic resonance
\item[NMR:] nuclear magnetic resonance
\item[QIP:] quantum information processing
\item[RF:] radio frequency
\item[SCROFULOUS:] short composite rotation for undoing length over and under shoot
\item[TROSY:] transverse relaxation optimized spectroscopy
\item[WALTZ:] wideband alternating-phase low-power technique for zero-residual-splitting
\end{description}

\end{document}